\documentclass[11pt]{article}
\pdfoutput=1
\usepackage{amssymb}
\usepackage[numbers,sort&compress]{natbib}
\usepackage{amsmath,amsfonts,graphicx,epsfig}
\usepackage{bm}
\usepackage{revsymb}

\def\bea{\begin{eqnarray}}
\def\eea{\end{eqnarray}}
\def\be{\begin{equation}}
\def\ee{\end{equation}}
\def\ba{\begin{array}}
\def\ea{\end{array}}

\newcommand{\ntot}{{\cal N}}
\newcommand{\calP}{{\cal P}}
\newcommand{\calR}{{\cal R}}

\begin{document}

\setlength\arraycolsep{2pt}

\renewcommand{\theequation}{\arabic{section}.\arabic{equation}}
\setcounter{page}{1}

\setlength\arraycolsep{2pt}

\begin{titlepage}

\rightline{\footnotesize{LPTENS-10-36}} \vspace{-0.2cm}
\rightline{\footnotesize{CPHT-RR-080.0910}} \vspace{-0.2cm}

\begin{center}

\vskip 1.0 cm

{\LARGE  \bf Features of heavy physics \\ in the CMB power spectrum}

\vskip 1.0cm

{\large
Ana Ach\'ucarro$^{a,b}$, Jinn-Ouk Gong$^{a}$\footnote{Present address: Theory Division, CERN, CH-1211 Geneva 23, Switzerland}, Sjoerd Hardeman$^{a}$
\\
Gonzalo A. Palma$^{c}$ and Subodh P. Patil$^{d,e}$
}

\vskip 0.5cm

{\it
$^{a}$Instituut-Lorentz Theoretical Physics, Universiteit Leiden \mbox{2333 CA Leiden, The Netherlands}
\\
$^{b}$Department of Theoretical Physics, University of the Basque Country
\mbox{UPV-EHU,P.O. Box 644, 48080 Bilbao, Spain}
\\
$^{c}$Physics Department, FCFM, Universidad de Chile \mbox{Blanco Encalada 2008, Santiago, Chile}
\\
$^{d}$Laboratoire de Physique Th\'eorique, Ecole Normale Sup\'erieure
\\
24 Rue Lhomond, Paris 75005, France
\\
$^{e}$Centre de Physique Th\'eorique, Ecole Polytechnique and CNRS
\\
Palaiseau cedex 91128, France
}

\vskip 1.5cm

\end{center}

\begin{abstract}

The computation of the primordial power spectrum in multi-field
inflation models requires us to correctly account for all relevant
interactions between adiabatic and non-adiabatic modes around and
after horizon crossing. One specific complication arises from
derivative interactions induced by the curvilinear trajectory of the
inflaton in a multi-dimensional field space. In this work we compute
the power spectrum in general multi-field models and show that
certain inflaton trajectories may lead to observationally
significant imprints of `heavy' physics in the primordial power
spectrum if the inflaton trajectory turns, that is, traverses a bend,
sufficiently fast (without interrupting slow roll), even in cases where the normal modes have masses approaching the cutoff of our theory. We emphasise that turning is defined with respect to
  the geodesics of the sigma model metric, irrespective of whether this
  is canonical or non-trivial.  The imprints generically take the form
  of damped superimposed oscillations on the power spectrum. In the
  particular case of two-field models, if one of the fields is
  sufficiently massive compared to the scale of inflation, we are able to
  compute an effective low energy theory for the adiabatic mode
  encapsulating certain relevant operators of the full multi-field
  dynamics. As expected, a particular characteristic of this effective
  theory is a modified speed of sound for the adiabatic mode which is
  a functional of the background inflaton trajectory and the turns
  traversed during inflation. Hence in addition, we expect
  non-Gaussian signatures directly related to the features imprinted
  in the power spectrum.

\end{abstract}

\end{titlepage}

\newpage


\section{Introduction}

Single field slow-roll inflation~\cite{Linde:1981mu, Albrecht:1982wi} successfully accounts for many of the observed properties of the cosmic microwave background (CMB), including the near scale invariance of the power spectrum of the primordial density fluctuations that seed the observed CMB anisotropies~\cite{Mukhanov:1981xt}. Although one could claim that a large subset of the simplest models of single field inflation remain perfectly compatible with current CMB precision measurements~\cite{Komatsu:2010fb, Larson:2010gs}, a direct and accurate reconstruction of the primordial spectrum from CMB data is still limited by various degeneracies in the priors and systematics adopted in our reconstructions~\cite{Bridle:2003sa, TocchiniValentini:2004ht, Mukherjee:2005dc}. It may certainly be the case that the CMB data implies the presence of various features in the primordial power spectrum other than the nearly scale invariant power law parametrization anticipated from the simplest single field slow roll models~\cite{Starobinsky:1992ts, Adams:2001vc, TocchiniValentini:2004ht, Gong:2005jr, Covi:2006ci, Hunt:2007dn, Ichiki:2009xs, Peiris:2009wp, Hamann:2009bz}. Upcoming data, such as that from the Planck satellite promises to provide new handles on the overall shape of the spectrum and, particularly in combination with other data sets, could help us determine the precise nature of any possible features in it. If present, such features could lead to quantitative new tests on the single field slow-roll paradigm~\cite{Kosowsky:1995aa, Copeland:1997mn} and constitute strong evidence in favour of the existence of additional degrees of freedom present during the evolution of density perturbations as the universe inflated.

One particularly compelling possibility which we wish to discuss in this report is of features in the spectrum generated by heavy (relative to the scale of inflation) degrees of freedom which do not necessarily decouple from the dynamics of the inflaton. Although the effects of massive degrees of freedom on the density perturbations are known to quickly dissipate during inflation, there are evidently still a number of contexts where features in the primordial spectrum due to heavy physics can survive. It is well understood, for example, that departures from a Bunch-Davies vacuum as the initial condition for the scalar fluctuations will result in oscillatory features in the power spectrum (see for example Refs.~\cite{Danielsson:2002kx, Jackson:2010cw}). Other contexts in which features are generated in the power spectrum involve particle production during brief intervals --much smaller than an $e$-fold-- as the universe inflates. Examples of this include those situations where a massive field coupled to the inflaton suddenly becomes massless at a specific point in field space~\cite{Chung:1999ve, Elgaroy:2003hp, Mathews:2004vu, Romano:2008rr, Barnaby:2009dd}. Here it is the transfer of energy out of the inflaton field and the subsequent backscatter of its fluctuations off the condensate of created quanta that can result in features in the power spectrum, as well as in its higher moments \cite{neil2, neil3}. Yet another context where such features have been shown to arise is in chain inflation, where instead of slowly rolling down a smooth continuous potential, the inflaton field gradually tunnels a succession of many vacua~\cite{Chialva:2008xh}.

The purpose of the present report is to demonstrate the existence of, and to understand the general conditions under which features in the power spectrum result in the context of inflation embedded in a multi-scalar field theory (see Refs.~\cite{Langlois:2008mn, Peterson:2010np, Cremonini:2010sv} for other recent discussions on this). For this we consider models of inflation where all of the scalar fields remain heavy except for one (the inflaton) which rolls slowly in some multi-dimensional potential. An effective field theory analysis tells us that in such scenarios, inflation should proceed in exactly the same way as in the single field case, with subleading corrections suppressed by the masses of the heavy scalar fields (see for example Ref.~\cite{Weinberg:2008hq}). In this framework it is easy to take for granted that a simple truncation of any available heavy degrees of freedom is  the same as having integrated them out. However, it can certainly be the case that the adiabatic approximation is no longer valid at some point along the inflaton trajectory (e.g. due to a ``sudden''  turn that mixes heavy and light directions), and higher derivative operators in the effective theory are no longer negligible even as inflation continues uninterrupted.

In various models of inflation in supergravity and string theory, the inflaton is embedded in a non-linear sigma model with typical field manifold curvatures of the string or Planck scale~\cite{GomezReino:2006wv, Covi:2008ea, Covi:2008cn}. In this type of scenario the inflaton traverses a curvilinear trajectory generating derivative interactions between the adiabatic and non-adiabatic modes\footnote{Here, by adiabatic mode we refer to the mode which fluctuates along the inflationary trajectory whereas non-adiabatic modes correspond to those whose fluctuations remain orthogonal to the trajectory. We will also frequently denote them as curvature and isocurvature modes in this report.}~\cite{Gordon:2000hv, GrootNibbelink:2000vx, GrootNibbelink:2001qt}. In this context, it is straightforward to appreciate heuristically that a sudden enough turn can excite modes normal to the trajectory and non-trivially modify the evolution of the adiabatic mode. We will see that the net effect of this trajectory will translate into damped oscillatory features superimposed on the power spectrum -- the transients after a sudden transfer of energy between the excited heavy modes and the much lighter inflaton mode, and the subsequent re-scattering of its perturbations off the condensate of heavy quanta that redshift in short time \footnote{We also note the investigations of \cite{xu1, xu2}, where inflation in a putative string landscape is modelled using a random potential. Here, the background inflaton effectively executes a random walk, resulting in features at all scales in the power spectrum.}.

\begin{figure}[tb]
\begin{center}
\includegraphics[scale=0.55]{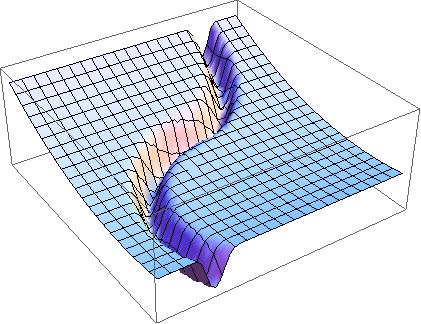}
\caption{\footnotesize A generic example of a potential where turns happens while one of the fields remain much heavier than the other.}
\label{figure-pot}
\end{center}
\end{figure}

A typical potential exhibiting such a curved trajectory is depicted in
Figure~\ref{figure-pot}. It can be appreciated that there is always a
heavy direction transverse to the loci of minima determining the
inflaton trajectory.  We should emphasise however that the focus of
this work is more general and that a curved trajectory in field space
is not exclusively due to the shape of the potential, but also depends
on the particular sigma model metric defining the scalar field
manifold: on a particular curve the two can be transformed into each
other by suitable field redefinitions. With this perspective, curved
trajectories appear in any situation where a mismatch exists between
the span of geodesics of the scalar field manifold and the actual
inflationary trajectory enforced by the scalar potential through the
equations of motion~\cite{Achucarro:2010jv}. The previously described
situation is in fact generic of realisations of inflation in the
context of string compactifications, where a large number of scalar
fields are expected to remain massive but with their vacuum
expectation values depending on the field value of the background
inflaton~\cite{BlancoPillado:2004ns, Lalak:2005hr, Conlon:2005jm,
  BlancoPillado:2006he, Simon:2006du, Bond:2006nc, de Carlos:2007dp,
  Lalak:2007vi, Grimm:2007hs, Linde:2007jn}.

Limits of certain cases we wish to study in this report have been
explored recently in seemingly different, but related contexts. In
Refs.~\cite{Chen:2009we, Chen:2009zp} for instance, the effects on
density perturbations due to a circular turn with constant curvature
in field space was explored within a two-field model. There, it was
concluded that such a turn could render non-Gaussian features in the
bispectrum but would not generate features in the power spectrum. In
another recent publication~\cite{Tolley:2009fg}, the effects of a
sigma model with non-canonical kinetic terms motivated by string
theory was explored within inflationary models where one of the fields
remained very massive. There, an effective theory was derived
describing the multi-field dynamics, characterised by having a speed
of sound for the fluctuations smaller than unity (and therefore
indicating the possible departure from Gaussianity of the CMB
temperature anisotropies). In the framework we are about to discuss,
both examples are just different faces of the same coin: while a
non-canonical sigma model metric can always be made locally flat along
a given trajectory generally generating contributions to the potential
with a curved locus of minima, it is also possible to find a field
redefinition which makes the loci of flat directions of the potential
look straight at the cost of introducing a non-canonical metric.

As we will shortly demonstrate, the parameter determining how relevant a local turn in the background inflaton trajectory is for the effective dynamics of the adiabatic mode is given by the departure from unity of the quantity $e^{\beta} = 1+4 \dot \phi_0^2 / (\kappa^2 M^2)$, where $\dot \phi_0$ is the speed of the inflaton background field, $\kappa$ is the radius of curvature of the curve in field space and $M$ is the mass of the direction normal to the trajectory. Keeping in mind that during slow-roll inflation, the inflaton velocity is given by $\dot \phi_0 = \sqrt{2 \epsilon} M_{\rm Pl} H$,  with $\epsilon$ being the usual slow roll parameter, it follows that $e^{\beta} = 1 + 8 \epsilon M_{\rm Pl}^2 H^2 / (\kappa^2 M^2)$. Thus, even with $M^2 \gg H^2$, if the radius of curvature describing the turn is small enough, significant imprints of heavy physics on the dynamics of the adiabatic mode can arise. More generally, whenever $e^\beta > 1$, some amount of particle creation takes place that backreacts on the dynamics of the adiabatic mode. Let us not forget that in addition to the scale invariance of the power spectrum, single field slow-roll inflation predicts that the observed CMB temperature anisotropies seeded by the curvature perturbation satisfy Gaussian statistics to a high degree of accuracy~\cite{Maldacena:2002vr}. Interestingly, in the class of models examined in this work, if the normal direction to the inflaton trajectory is sufficiently massive ($M^2 \gg H^2$), it is possible to compute an effective action for the adiabatic mode capturing the relevant operators of the full multi-field dynamics. This effective theory has the characteristic that the adiabatic mode propagates with a speed of sound given by
\begin{equation}
c_s^2 = e^{-\beta}
\end{equation}
and therefore becomes a functional of (the curvature of) the trajectory traversed by the
inflaton~\cite{Achucarro:2010jv}. Interestingly, this result has as a
special case the particular context of Ref.~\cite{Tolley:2009fg} and
indicates the presence of non-Gaussian signatures correlated with
features in the power spectrum.\footnote{In the course of preparing
  this manuscript, we note with interest the results of \cite{Cremonini-arxiv}, where their parametrization of the
  non-decoupling parameter of the isocurvature directions $\xi$
  relates as a specific realisation of our analysis. This is easiest
  seen through comparing expressions (23) in Ref~\cite{Cremonini-arxiv}
  with (\ref{betadef1}) or (\ref{betadef2}) here. We emphasise that
  the non-decoupling effect, ascribed in these two papers to the
  non-trivial sigma model metric, is simply due to the non-geodesic
  nature of the trajectory obtained by putting the heavy field at its
  local minimum.}

We have organised this paper in the following way. In
Section~\ref{sec2: setup} we present the general setup and the
notations used throughout this work. There, we will emphasise the need
for using a geometric perspective to describe the evolution of the
homogeneous background.  Then in Section~\ref{sec3: perturbations} we
proceed to examine the perturbations of the fields around a time
dependent background, and consider their quantisation and provide
general formulas for the power spectrum.  Our formalism allows us to
consider situations beyond the regime of applicability of existing
methods, such as trajectories with fast, sudden turns (regardless of
whether the sigma model metric is canonical or non-canonical), and any
other situations in which the masses in the orthogonal direction are
changing relatively fast along the trajectory while still remaining
much heavier than $H^2$. In Section~\ref{two-field-power} we apply the
previous results to the particular case of two-field models. We also
deduce an effective theory valid in the case where the field normal to
the trajectory remains heavy and compare the spectrum of this
effective, single-field theory with those found in the full,
multi-field computation of the spectrum. Then, in
Section~\ref{sec-results} we examine in detail the evolution of
adiabatic and non-adiabatic modes along the curved inflaton trajectory
and compute the power spectrum for various cases. Finally, in
Section~\ref{sec: conclusions} we provide our concluding remarks.


\section{Basic considerations}
\setcounter{equation}{0}
\label{sec2: setup}

Let us start our study by recalling some of the basic aspects of multi-field inflation and by introducing the notations and conventions that will be used throughout this work. Our starting point is to assume the following effective four dimensional action consisting of gravity and a set of  $\ntot$ scalar fields $\phi^a$:
\begin{equation}
S = \int \!\! \sqrt{-g} \, d^4 x \left[  \frac{M_{\rm Pl}^2}{2 } R - \frac{1}{2} \gamma_{a b} g^{\mu \nu} \partial_{\mu} \phi^a  \partial_{\nu} \phi^b - V(\phi) \right] \, .
\label{sec-1:act}
\end{equation}
Here $R$ denotes the Ricci scalar constructed out of the spacetime metric $g_{\mu \nu}$ with determinant $g$. Additionally, $\phi^a$ ($a = 1, \cdots  \ntot$) denotes a set of scalar fields spanning a scalar manifold $\mathcal{M}$ of dimension $\ntot$, equipped with a scalar metric $\gamma_{a b}$. The scalar fields may be thought of as coordinates on $\mathcal{M}$ with Christoffel symbols given by
\begin{equation}
\Gamma^{a}_{b c} = \frac{1}{2} \gamma^{a d} \left( \partial_b \gamma_{d c} + \partial_c \gamma_{b d} - \partial_d \gamma_{b c} \right) \, ,
\end{equation}
where $\partial_a$ are partial derivatives with respect to the scalar fields $\phi^a$. In terms of these, the Riemann tensor associated with $\mathcal{M}$ is given by
\begin{equation}
\mathbb{R}^a{}_{b c d} = \partial_c \Gamma^a_{b d} - \partial_d \Gamma^a_{b c} + \Gamma^a_{c e} \Gamma^e_{d b} -   \Gamma^a_{d e} \Gamma^e_{c b} \, .
\end{equation}
It is also possible to define the Ricci tensor as $\mathbb{R}_{a b} =  \mathbb{R}^c{}_{a c b}$ and the Ricci scalar $\mathbb{R} = \gamma^{a b} \mathbb{R}_{a b}$. We shall be careful to distinguish geometrical quantities related to the four dimensional spacetime and the $\ntot$-dimensional abstract manifold $\mathcal{M}$. We should keep in mind that, typically, there will be an energy scale $\Lambda_{\mathcal M}$ associated to the curvature of $\mathcal{M}$, and hence, fixing the typical mass scale of the Ricci scalar as $\mathbb{R} \sim \Lambda_{\mathcal M}^{-2}$. In many concrete situations, such as the modular sector of string compactifications, the scale $\Lambda_{\mathcal M}$ corresponds to the Planck mass $M_{\rm Pl}$. The equations of motion for the scalar fields are given by
\be
\Box \phi^a + \Gamma^a_{b c} g^{\mu \nu}  \partial_{\mu} \phi^b \partial_{\nu} \phi^c = V^a \,, \label{scalar-equation-1}
\ee
where $V^a \equiv \gamma^{a b} \partial_b V$. In what follows we discuss in detail the homogeneous solutions $\phi^a = \phi^a_0(t)$ to these equations where the scalar fields depend only on time. The discussion in the sequel follows closely the analysis in \cite{GrootNibbelink:2000vx, GrootNibbelink:2001qt}.

\subsection{Background solution}
\label{section-bacground-solution}

We look for background solutions by assuming that all the scalar fields are time dependent $\phi^a = \phi^a_0(t)$, and that spacetime consists of a flat Friedmann-Robertson-Walker (FRW) geometry of the form
\begin{equation}
ds^2 = - dt^2 + a^2(t)  \delta_{ij}dx^idx^j \, .
\end{equation}
Later on we will also work in conformal time $\tau$, defined through the relation $d t = a \, d\tau$. In this background, the equation of motion (\ref{scalar-equation-1}) describing the evolution of the scalar fields is given by
\begin{equation}
\frac{D}{dt} \dot \phi_0^a + 3 H \dot \phi_0^a + V^a = 0 \, ,
\label{scal-eq-1}
\end{equation}
where $H = \dot a / a$ is the Hubble parameter characterising the expansion rate of spatial slices, and where we have also introduced the convenient notation $D X^a = d X^a + \Gamma^a_{b c} X^b d \phi_0^c$. On the other hand, the Friedmann equation describing the evolution of the scale factor in terms of the scalar field energy density is given by
\begin{equation}
H^2 = \frac{1}{3 M_{\rm Pl}^2} \left( \frac{1}{2} \dot \phi_0^2 + V  \right) \, ,
\label{friedmann-eq-1}
\end{equation}
where $\dot \phi_0^2 \equiv \gamma_{a b} \dot \phi_0^a  \dot \phi_0^b$. Notice that $ \dot \phi_0 $ corresponds to the rate of change of the scalar field vacuum expectation value along the trajectory followed by the background fields. It is also convenient to recall the following equation describing the variation of $H$
\begin{equation}
\dot H = - \frac{\dot \phi_0^2}{2 M_{\rm Pl}^2} \, ,
\label{acceleration-eq-1}
\end{equation}
which may be deduced by combining (\ref{scal-eq-1}) and (\ref{friedmann-eq-1}). By specifying the metric $\gamma_{a b}$ and the scalar potential $V$, these equations can be solved to obtain the curved trajectory in $\mathcal{M}$ followed by the scalar fields. To discuss several features of this trajectory without explicitly solving the previous equations, it is useful to define unit vectors $T^a$ and $N^a$ distinguishing tangent and normal directions to the trajectory respectively, in such a way that $T^a N_a = 0$. These are defined as
\begin{align}
T^a \equiv & \frac{\dot \phi_0^a}{\dot \phi_0} \, , \nonumber
\\
N^a \equiv & s_N(t) \left( \gamma_{b c} \frac{D T^b}{dt} \frac{D T^c}{dt} \right)^{- 1/2} \frac{D T^a}{dt} \, ,
\label{eq:def-t-n}
\end{align}
where $s_N (t) = \pm 1$, denoting the orientation of $N^a$ with
respect to the vector $D T^a/dt$. That is, if $s_N (t) = +1$ then
$N^a$ is pointing in the same direction as $D T^a/dt$, whereas if $s_N
(t) =-1$ then $N^a$ is pointing in the opposite direction.  Due to the
presence of the square root, it is clear that $N^a$ is only well
defined at intervals where $D T^a/dt \neq 0$. However, since $D
T^a/dt$ may become zero at finite values of $t$, we allow $s_N (t)$
to flip signs each time this happens in such a way that both $N^a$ and $DT^a/dt$
remain a continuous function of $t$. This implies that the sign of $s_N$ may be chosen conventionally at some initial time $t_i$, but from then on it is subject to the equations of motion respected by the background.%
%
\footnote{We are assuming here that the background solutions $\phi^a = \phi_0^a(t)$ are analytic functions of time, and therefore we disregard any situation where this procedure cannot be performed.}    
In the particular case where $\mathcal{M}$ is two dimensional, the
presence of $s_N(t)$ in (\ref{eq:def-t-n}) is sufficient for $N^a$ to
have a fixed orientation with respect to $T^a$ (either left-handed or
right-handed). This will become particularly useful when we examine
two dimensional models in Section \ref{two-field-power}.

Observe that the tangent vector $T^a$ offers an alternative way of defining the total time derivative $D/dt$ along the trajectory followed by the scalar fields. This is:
\begin{equation}
\frac{D}{ dt } \equiv  \dot \phi_0 T^a \nabla_a  =   \dot \phi_0  \nabla_\phi \, .
\label{Dt-td}
\end{equation}
Now, taking a total time derivative to $T^a$, we may  use the  equation of motion (\ref{scal-eq-1})  to write
\begin{equation}
 \frac{D T^a}{dt} = - \frac{\ddot \phi_0}{\dot \phi_0} T^a - \frac{1}{\dot \phi_0} \left( 3 H \dot \phi_0^a + V^a \right) \, .
\end{equation}
Then, by projecting this equation along the two orthogonal directions $T^a$ and $N^a$, we obtain the following two independent equations
\begin{align}
\ddot \phi_0 + 3 H \dot \phi_0 + V_{\phi} = 0 \, ,
\label{eq-mot-sigma}
\\
\frac{D T^a}{dt} = - \frac{V_N}{\dot \phi_0} N^a \, ,
\label{eq-mot-t-n-1}
\end{align}
where we have defined $V_{\phi} \equiv T^a V_a$ and $V_{N} \equiv N^a V_a$ to be the projections of $V_a = \partial_a V$ along the tangent and normal directions respectively. It is not difficult to verify that $V_{a}$ lies entirely along a space spanned by $T^a$ and $N^a$. That is, we are allowed to write $V_{a} \equiv V_{\phi} T_{a} + V_{N} N_{a}$. To anticipate the study of inflation within the present setup, it is useful to define the following dimensionless quantities:
\begin{align}
\epsilon \equiv & - \frac{\dot H}{H^2} = \frac{\dot \phi_0^2}{2 M_{\rm Pl}^2 H^2} \, ,
\\
\eta^a \equiv & - \frac{1}{H \dot \phi_0}  \frac{D \dot \phi_0^a}{dt} \, .
\label{def-slow-roll-parameters}
\end{align}
We will not assume that these parameters are small until much later, where inflation is studied in the slow-roll regime (see Section~\ref{section: slow-roll}). Similarly to the case of $V_a$, the vector $\eta^a$ may be decomposed entirely in terms of $T^a$ and $N^a$ as
\begin{align}
\eta^a = & \eta_{||} T^a + \eta_{\bot} N^a \, , \label{deff-slow-roll-parameters-1}
\\
\eta_{||} \equiv & -\frac{\ddot \phi_0}{H \dot \phi_0} \, ,  \label{deff-slow-roll-parameters-2}
\\
\eta_{\bot} \equiv & \frac{V_N}{\dot \phi_0 H} \, ,  \label{deff-slow-roll-parameters-3}
\end{align}
where we have used (\ref{scal-eq-1}) to simplify a few expressions. Observe that $\eta_{\bot}$ is directly related to the rate of change of the tangent unit vector $T^a$, since (\ref{eq-mot-t-n-1}) can be written as
\begin{align}
 \frac{D T^a}{dt}  = - H \eta_{\bot} N^a \, .
 \label{D-H-eta-bot}
\end{align}
Comparison with (\ref{eq:def-t-n}) shows that 
${\rm sign}(\eta_{\bot}) = -s_N$. This is one of our main reasons for
having introduced $s_N(t)$ in (\ref{eq:def-t-n}): it allows us to keep
$\eta_{\bot}$ continuous and avoid some unnecessary difficulties
encountered in the definition of isocurvature modes.\footnote{In
  Ref.~\cite{Peterson:2010np} for instance, a similar parameter
  $\eta_{\bot}$ is introduced but with a fixed sign, which if the `slow turn' approximation were to be violated (a regime we are mainly interested in), would produce a numerical overshoot in the evolution of curvature and isocurvature perturbations that has to be taken into account, and is naturally accounted for by our definition.}

\begin{figure}[tb]
\begin{center}
\includegraphics[scale=0.6]{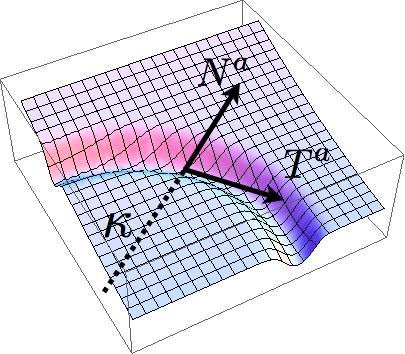}
\caption{\footnotesize The figure shows schematically the relation between the tangent vector $T^a$, the normal vector $N^a$ and the radius of curvature $\kappa$.}
\label{fig-N-T-k}
\end{center}
\end{figure}

Moving on with this discussion, we can relate $\eta_{\bot}$ to the radius of curvature $\kappa$ characterising the bending of the trajectory followed by the scalar fields. To do so, let us recall that given a curve $\gamma(\phi_0)$ in field space parameterised by $d \phi_0 = \dot \phi_0 dt$, we may define the radius of curvature $\kappa$ associated to that curve through the following relation:
\begin{equation}
 \frac{1}{\kappa} =   \left( \gamma_{b c} \frac{D T^b}{d \phi_0} \frac{D T^c}{d \phi_0} \right)^{1/2} \, .
 \label{def-kappa}
\end{equation}
Here $\kappa$ stands for the radius of curvature in the scalar manifold $\mathcal M$ spanned by the $\phi^a$ fields, and therefore it has dimension of mass. Figure~\ref{fig-N-T-k} shows the relation between the pair of vectors $T^a$, $N^a$ and the radius of curvature $\kappa$. Using (\ref{Dt-td}) and comparing the last two equations we find the following relation between $\kappa$ and $\eta_{\bot}$:
\begin{equation}
\kappa^{-1} =  \frac{ H  | \eta_{\bot} |}{ \dot \phi_0 } \, .
\label{eta-bot-kappa}
\end{equation}
By definition any geodesic curve $\gamma(\phi_0)$ in $\mathcal{M}$ satisfies the relation $D \dot \phi^a /d t  \propto  \dot \phi^a$, which corresponds to the case $\kappa^{-1} = 0$, or alternatively, to the case $\eta_{\bot} = 0$. Thus, we see that the dimensionless parameter $\eta_{\bot}$ is a useful quantity that parameterises the bending of the inflationary trajectory with respect to geodesics in $\mathcal{M}$. It is interesting to rewrite the previous relation by replacing $\dot \phi_0 = \sqrt{2 \epsilon} H M_{\rm Pl}$ coming from the definition of $\epsilon$ presented in (\ref{def-slow-roll-parameters}). One obtains:
\begin{equation}
| \eta_{\bot} | = \sqrt{2 \epsilon} \frac{M_{\rm Pl}}{\kappa} \, .
\label{eta-bot-kappa-2}
\end{equation}
Then, if the radius of curvature is such that $\kappa \ll M_{\rm Pl}$, one already sees that $\eta_{\bot}^2 \gg 2 \epsilon$. We shall come back to this result later when we study curved trajectories in the slow-roll regime $\epsilon \ll 1$. To continue, we may further characterise the variation of $N^a$ as:
\begin{equation}
 \frac{D  N^a}{dt} = H \eta_{\bot} T^a + \frac{1}{H \eta_{\bot}}  P^{a b} \nabla_{\phi} V_b \, ,
 \label{D-n}
\end{equation}
where we have defined the projector tensor $P^{ab} \equiv \gamma^{a b} - T^a T^b - N^a N^b$ along the space orthogonal to the subspace spanned by the unit vectors $T^a$ and $N^a$. That is, $P_{a b} N^b = 0$ and $P_{a b} T^b = 0$.  Details on how to obtain the previous relation can be found in Appendix~\ref{background-app}. Observe that in the particular case where $\mathcal{M}$ is two dimensional, one has $\gamma_{a b} = T_a T_b + N_a N_b$ and therefore $P_{a b} = 0$ identically.

To finish this section, let us state some useful relations that will be used throughout the rest of this work. First, by using the definitions for $\epsilon$ and $\eta_{||}$ in (\ref{def-slow-roll-parameters}), we may rewrite the background equations (\ref{friedmann-eq-1}) and (\ref{eq-mot-sigma}) respectively as:
\begin{align}
 3 - \epsilon =& \frac{V}{M_{\rm Pl}^2H^2} \, ,
 \label{non-slow-1}
 \\
 3 - \eta_{||} =& - \frac{ V_{\phi} }{ \dot \phi_0 H } \, .
 \label{non-slow-2}
\end{align}
With the help of (\ref{def-slow-roll-parameters}) these two relations may be put together to yield:
\begin{equation}
\epsilon =  \frac{M_{\rm Pl}^2}{2} \left( \frac{   V_{\phi} }{ V } \right)^2 \left( \frac{3 - \epsilon}{3 - \eta_{||} } \right)^2 \, .
\label{first-slow-cond}
\end{equation}
Next, by deriving (\ref{eq-mot-sigma}) with respect to time and using the definitions for $\epsilon$ and $\eta_{||}$, we deduce
\begin{align}
3 (\epsilon + \eta_{||}) = & M_{\rm Pl}^{2}\frac{\nabla_{\phi} V_{\phi} }{ V } (3 - \epsilon)+ \xi_{||} \eta_{||} \, ,
\label{second-slow-cond}
\\
\xi_{||} \equiv & - \frac{1}{H \ddot \phi_0}
 \dddot\phi_0 .
\end{align}
Both (\ref{first-slow-cond}) and (\ref{second-slow-cond}) are exact equations linking the evolution of background quantities with the scalar potential $V$. It may be already noticed that if $\epsilon$, $\eta_{||}$ and $\xi_{||}$ are all much smaller than unity, then we obtain the usual relations for the slow roll parameters $\epsilon$ and $\eta_{||}$ in terms of derivatives of the potential:
\begin{align}
\epsilon \approx & \frac{M_{\rm Pl}^2}{2} \left( \frac{   V_{\phi} }{ V } \right)^2 \, ,
\\
\epsilon + \eta_{||} \approx & M_{\rm Pl}^{2}\frac{\nabla_{\phi} V_{\phi} }{ V }  \, .
\end{align}
We shall come back to these relations later, when we consider the evolution of the background in the slow roll regime.


\section{Perturbation theory}
\setcounter{equation}{0}
\label{sec3: perturbations}

The  notation introduced in the previous section provides a useful tool to analyse perturbations $\delta \phi^a$ about the background solution $\phi^a = \phi^a_0(t)$ by decomposing them into parallel and normal components with respect to the inflaton trajectory. In what follows we proceed to study the evolution and quantisation of these perturbations. First, we consider scalar field perturbations by expanding about the background $\phi^a(t,{\bm x}) = \phi_0^a(t) + \delta \phi^a(t,{\bm x})$. It is well known that the equations of motion for the perturbed fields can be cast entirely in terms of the gauge-invariant Mukhanov-Sasaki variables~\cite{Sasaki:1986hm, Mukhanov:1988jd}
\begin{equation}
Q^a \equiv \delta \phi^a + \frac{\dot \phi^a}{H} \psi \, ,
\end{equation}
where  $\psi$ is the curvature perturbation of the spatial metric.
The equations of motion for these fields are found to be~\cite{Sasaki:1995aw}
\begin{equation}
\frac{D^2 Q^a}{dt^2} + 3 H \frac{D Q^a}{dt} - \frac{\nabla^2}{a^2} Q^a + C^a{}_b Q^b = 0 \, , \label{pert-eq-1}
\end{equation}
where  $\nabla^2 \equiv \delta^{ij}\partial_i\partial_j$ is the spatial Laplacian
and where we have defined the tensor $C^a{}_b$ as:
\begin{equation}
C^a{}_b \equiv \nabla_b V^a - \dot \phi_0^2 \mathbb{R}^a{}_{c d b} T^c T^d + 2\epsilon \frac{H}{  \dot \phi_0}  \left( T^a V_b + T_b V^a \right)  + 2 \epsilon (3 - \epsilon)  H^2 T^a T_b \, . \label{definition-C}
\end{equation}
We notice here that $C_{a b} = \gamma_{a c} C^c{}_b$ is symmetric. It is convenient to rewrite the set of  equations (\ref{pert-eq-1}) in terms of perturbations orthogonal to each other. With this in mind, we introduce a complete set of vielbeins $e^I_a = e^I_a(t)$ and work with the following quantities:
\begin{equation}
Q^I (t,{\bm x}) \equiv e^I_a (t) Q^a (t,{\bm x}) \, .
\label{Q-Q}
\end{equation}
The $a$-index labels the abstract scalar manifold $\mathcal M$ whereas the $I$-index labels a local orthogonal frame moving along the inflationary trajectory.
Recall that vielbeins are defined to satisfy the basic relations $e^I_a e^J_b \gamma^{a b} = \delta^{I J}$ and $e^I_a e^J_b \delta_{I J} = \gamma_{a b}$. From these relations one deduces the identities
\begin{align}
e^{I}_a \frac{D}{dt} e^a_J = & - e^a_J \frac{D}{dt} e^I_a \, ,
\\
e^{a}_I \frac{D}{dt} e_b^I = & - e_b^I \frac{D}{dt} e_I^a \, ,
\end{align}
from which it is possible to read
\begin{align}
\dot Q^I =& e^I_a \frac{D Q^a}{dt} - Y^I{}_J Q^J \, ,
\label{Q-dot-1}
\\
\ddot Q^I =& e^I_a \frac{D^2 Q^a}{dt^2} - 2 Y^I{}_J \dot Q^J - \left( Y^I{}_K Y^K{}_J + \dot Y^I{}_J \right) Q^J \, ,
\label{Q-dot-2}
\end{align}
where we have defined the antisymmetric matrix $Y_{I J} = - Y_{J I}$ as:
\begin{equation}
 Y^I{}_J = e^I_a  \frac{D e^a_J}{dt} \, .
 \label{def-Z}
\end{equation}
Before writing down the equations of motion respected by the fields $Q^I$, it is useful to notice that the matrix $Y_{IJ}$ allows us to define a new covariant derivative $\mathcal{D}/dt$ acting on quantities such as $Q^I$ labelled with the $I$-index in the following way\footnote{It may be noticed that we can write $Y^{I}{}_J = \left( e^I_a \partial_b e^a_J + e^I_a \Gamma^{a}_{b c} e^c_J \right) \dot \phi_0^b = \omega_b{}^I{}_J \dot \phi_0^b$ where $\omega_b{}^I{}_J$ are the usual spin connections for non-coordinate basis, hence justifying the definition of the new covariant derivative of (\ref{def-D-X-2}).}:
\begin{equation}
\frac{\mathcal D}{d t} Q^I  \equiv  \dot Q^I + Y^{I}{}_J Q^J \, .
\label{def-D-X-2}
\end{equation}
This definition allows us to rearrange (\ref{Q-dot-1}) and~(\ref{Q-dot-2}) and simply write
\begin{align}
\frac{\mathcal{D} Q^I}{d t}  = & e^I_a \frac{D Q^a}{dt} \, ,
\\
\frac{\mathcal{D}^2 Q^I}{d t^2}  = & e^I_a \frac{D^2 Q^a}{dt^2} \, .
\end{align}
Thus, the equations of motion for the perturbations in the new basis become
\begin{equation}
\frac{\mathcal{D}^2 Q^I}{d t^2} + 3H \frac{\mathcal{D} Q^I}{d t} - \frac{\nabla^2}{a^2} Q^I + C^I{}_J Q^J = 0 \, ,
\end{equation}
where $C_{IJ}  \equiv e_{I a} e^b_J C^a{}_{b}$. To deal with the above set of equations, it is convenient to take one last step in simplifying them and rewrite them in terms of conformal time $d \tau = dt/a$, and a new set of perturbations $v^I \equiv a Q^I$. These redefinitions induce a re-scaling  of the covariant derivative (\ref{def-D-X-2}) in the form ${\mathcal D}/{d \tau} = a {\mathcal D}/{d t}$, from where we are allowed to write:
\begin{equation}
\frac{\mathcal{D} v^I}{d\tau} = \frac{d v^I}{d \tau}  + Z^{I}{}_{J} v^{J} \, ,
\label{def-D-X-2-new}
\end{equation}
where $Z_{IJ} = a Y_{I J}$. With this notation at hand,  the equations of motion for the $v^I$-perturbations are found to be
\begin{equation}
\frac{ \mathcal{D}^{2} v^I}{d \tau^2} -  \nabla^2 v^I+  \Omega^{I}{}_{J} v^J = 0 \, ,
\label{eom-cov-1}
\end{equation}
where $\Omega_{I J} =  -  a^2 H^2 (2 - \epsilon)  \delta_{I J} + a^2   C_{I J}$ and we have used the definition of  $\epsilon$ to write $a''/a = a^2 H^2 (2 - \epsilon)$. For completeness, we notice that the previous equations of motion may be derived from the following action:
\begin{equation}
S = \frac{1}{2} \int d \tau d^3 x \left[   \sum_I \left( \frac{\mathcal{D} v^I}{d \tau} \right)^2 - \sum_I ( \nabla v^I  )^2 - \Omega_{I J} v^I v^J \right] \, ,
\label{action-for-v-fields}
\end{equation}
which can be alternatively deduced directly from the initial action (\ref{sec-1:act}) by considering all of the field redefinitions introduced in the present discussion.

The set of equations (\ref{eom-cov-1}) contains several non-trivial features. First, notice that the covariant derivative ${\mathcal D}/{d \tau}$ implies the existence of non-trivial couplings affecting the kinetic term of each field $v^I$. By the same token, under general circumstances  the symmetric matrix $\Omega_{IJ}$ does not remain diagonal at all times. In fact, as we are about to see in the next section, it is possible to choose to write this theory either in a frame where the  $\ntot$ scalar fields are canonical (and therefore without non-trivial couplings in the kinetic term), or either in a frame where $\Omega_{IJ}$ remains diagonal, but (in general) not both at the same time.


\subsection{Canonical frame}
\label{canonical-frame}

Observe that by introducing the vielbeins $e^{I}_a$ in the previous section, we have not specified any alignment of the moving frame.
In fact, given an arbitrary frame, characterised by the set $e^{I}_a$, it is always possible to find a canonical frame where the scalar field perturbations acquire canonical kinetic terms in the action. To find it, let us  introduce a new set of fields $u^I$ defined out of the original fields  $v^I$ in the following way:
\begin{equation}
v^I (\tau, {\bm x}) = R^I{}_{J} (\tau , \tau_i ) u^J (\tau, {\bm x}) \, ,
\label{rel-u-v-1}
\end{equation}
where $R^I{}_{J}(\tau, \tau_i )$ is an invertible matrix defined to satisfy the following first order differential equation:
\begin{equation}
\frac{d}{d \tau} R^{I}{}_{J} = - Z^{I}{}_{K} R^{K}{}_{J} \, ,
\label{R-def}
\end{equation}
with the boundary condition $R^I{}_{J}(\tau_i , \tau_i ) = \delta^{I}{}_J$ set at some given initial
time $\tau_i$. Let us additionally define a new matrix $S^{I}{}_{J}$ to be the inverse of  $R^{I}{}_{J}$, i.e.
$S^{I}{}_{K} R^{K}{}_{J} = R^{I}{}_{K} S^{K}{}_{J} = \delta^{I}{}_{J}$. Then, it is possible to see that $S^{I}{}_{J}$ satisfies the following similar equation
\begin{equation}
\frac{d}{d \tau} S_{I}{}^{J} = - Z^{J}{}_{K} S_{I}{}^{K} \, ,
\label{S-def}
\end{equation}
where we used the fact that $Z_{I J} = - Z_{J I}$.
Since both solutions to (\ref{R-def}) and (\ref{S-def}) are unique, then the previous equation tells us that $S_{IJ} = R_{J I}$, that is, $S$ corresponds to $R^T$ the transpose of $R$. This means that for a fixed time $\tau$, $R_{IJ}(\tau, \tau_i)$ is an element of the orthogonal group O$( \ntot)$, the group of matrices $R$ satisfying $R R^T = \openone$. The solution to (\ref{R-def}) is well known, and may be symbolically written as
\begin{equation}
R (\tau, \tau_i) =
 \openone
+ \sum_{n=1}^{\infty} \frac{(-1)^n}{n!} \!\! \int_{\tau_i}^\tau  \mathcal{T} \left[ Z (\tau_1) \cdots Z(\tau_n) \right] d^n \tau
= \mathcal{T} \exp \left[ -  \int_{\tau_i}^\tau \!\!\! d \tau \, Z(\tau) \right] \, ,
\end{equation}
where $\mathcal{T}$ stands for the usual time ordering symbol, that is  $\mathcal{T} [Z (\tau_1) Z (\tau_2) \cdots Z(\tau_n)]$ corresponds to the product of $n$ matrices $Z (\tau_i)$ for which $\tau_1 \ge \tau_2 \ge \cdots \ge \tau_n$. Coming back to the  $u^I$-fields, it is possible to see now that, by virtue of (\ref{R-def}) one has:
\begin{align}
\frac{\mathcal{D} v^I}{d \tau} = & R^{I}{}_{J} \frac{d u^{J}}{d \tau} \, ,
\\
\frac{\mathcal{D}^2 v^I}{d \tau^2} = & R^{I}{}_{J} \frac{d^2 u^{J}}{d \tau ^2} \, .
\end{align}
Inserting these relations back into the equation of motion (\ref{eom-cov-1}) we obtain the following equation of motion for the $u^I$-fields:
\begin{equation}
\frac{ d^{2} u^I}{d \tau^2}  - \nabla^2 u^I +  \left[  R^T (\tau) \, \Omega \, R (\tau) \right]^{I}{}_{J} u^J = 0 \, .
\label{eom-u-cov-1}
\end{equation}
Additionally, it is possible to show that the action (\ref{action-for-v-fields}) is now given by
\begin{equation}
S = \frac{1}{2} \int d \tau d^3 x \left\{ \sum_I \left( \frac{d u^I}{d\tau} \right)^2  -  \left( \nabla u^I \right)^2
- \left[  R^T (\tau) \Omega R(\tau) \right]_{I J} u^I u^J \right\} \, .
\label{canonical-action}
\end{equation}
Thus, we see that the fields $u^I$ correspond to the canonical fields in the usual sense. This result shows, just as we have stated, that it is always possible to find a frame where the perturbations become canonical, but at the cost of having a mass matrix $\left[  R^T (\tau) \, \Omega \, R (\tau) \right]_{I J}$ with non-diagonal entries which are changing continuously in time. Another way to put it is that, while both $ R^T (\tau) \, \Omega \, R (\tau) $ and $\Omega$ share the same eigenvalues, as long as $R (\tau)$ varies in time, their associated eigenvectors will not remain aligned.  To finish, let us notice that by construction, at the initial time $\tau_i$, the canonical fields $u^I$ and the original fields $v^I$ coincide $u^I (\tau_i) = v^I (\tau_i)$. However, it is always possible to redefine a new set of canonical fields by performing an orthogonal transformation of the fields.


\subsection{Quantisation and initial conditions}
\label{quant-and-initial}

Having the canonical frame at hand, we may now quantise the system in the standard way. Starting from the action (\ref{canonical-action}) it is possible to see that  the canonical coordinate fields are given by $u^I$ whereas the canonical momentum is given by $\Pi_u^I  = { d u^I }/{d \tau}$. To quantise the system, we demand this pair to satisfy the commutation relation
\begin{equation}
\left[ u^I (\tau,{\bm x}) , \Pi_u^J (\tau,{\bm y}) \right] = i \delta^{I J} \delta^{(3)} ({\bm x} - {\bm y}) \, ,
\label{commutation-0}
\end{equation}
otherwise zero. With the help of the $R$ transformation introduced in (\ref{rel-u-v-1}) we can rewrite this commutation relation to be valid in an arbitrary moving frame. More precisely, we observe here that we are allowed to define a new pair  of fields $v^I$ and $\Pi^I_v$ given by
\begin{align}
v^I = & R^{I}{}_{J} u^J \, ,
\\
\Pi_{v }^{I}  \equiv & \frac{\mathcal D}{d \tau} v^I  =  R^{I}{}_{J} (\tau, \tau_i)  \Pi_{u}^{J} \, .
\end{align}
From (\ref{commutation-0}), this new pair is found to satisfy similar commutation relations:
\begin{equation}
\left[ v^I (\tau,{\bm x}) , \Pi_v^J (\tau,{\bm y}) \right] = i \delta^{I J} \delta^{(3)} ({\bm x} - {\bm y}) \, .
\label{commutation-1}
\end{equation}
Following the convention, it is now possible to obtain an explicit expression for $v^I({\rm x},\tau)$ in terms of creation and annihilation operators.\footnote{From this point on, we continue working with the more general $v^I$-fields instead of the canonical $u^I$-fields. Nevertheless, we emphasise that the $u^I$-fields allowed us to find the correct quantisation prescription for the $v^I$-fields, which otherwise would not have been properly justified.} For this, let us write $v^I({\bf x} ,\tau)$ as a sum of Fourier modes:
\begin{align}
v^I (\tau,{\bm x}) = & \int \frac{d^3k}{(2\pi)^{3/2}} e^{i{\bm k}\cdot{\bm x}} v^I({\bm k},\tau)
\nonumber\\
= & \int \frac{d^3k}{(2\pi)^{3/2}} e^{i{\bm k}\cdot{\bm x}} \sum_{\alpha} \left[ v_{\alpha }^I (k,\tau) \, a_\alpha ({\bm k}) + v_{\alpha }^{I *}(k, \tau) \, a^{\dag}_\alpha (-{\bm k}) \right] \, .
\label{v-fourier}
\end{align}
In writing the previous expression we have anticipated the need of expressing the fields $v^I (\tau,{\bm x})$ as a linear combination of $\ntot$ time-independent creation and annihilation operators $a^{\dag}_\alpha ({\bm k})$ and $a_\alpha ({\bm k})$ respectively, with $\alpha = 1, \cdots  \ntot$. These operators are required to satisfy the usual relations
\begin{equation}
\left[ a_\alpha ({\bm k}) , a_\beta^{\dag}({\bm q}) \right] = \delta_{\alpha \beta}  \delta^{(3)} ({\bm k} - {\bm q}) ,
\end{equation}
 otherwise zero. This set of operators defines the vacuum $| 0 \rangle$ of the theory by their action $a_{\alpha} ({\bm k}) | 0 \rangle = 0$. Since the operators  $a^{\dag}_\alpha ({\bm k})$ and $a_\alpha ({\bm k})$, for different values of $\alpha$, are taken to be linearly independent, then the time-dependent coefficients $v_{\alpha }^I (k, \tau)$ appearing in front of them in (\ref{v-fourier}) must satisfy the equation of motion:\footnote{It is crucial to appreciate that the Greek indices $\alpha$ label scalar quantum modes and not directions in field space, as capital Latin indices do. Different $\alpha$-modes may contribute to the same fluctuation along a given direction $I$. The quantities linking these two different abstract spaces are the mode functions $v_{\alpha }^I (k, \tau)$ whose time evolution is dictated by (\ref{eq-motion-v}). A similar scheme to quantise a coupled multi-scalar field system may be found in Ref.~\cite{Nilles:2001fg}.}
\begin{equation}
\frac{\mathcal{D}^2}{d \tau^2} v_{\alpha}^{I} (k,\tau) + k^2 v_{\alpha}^{I} (k,\tau) + \Omega^{I}{}_{J} v_{\alpha}^{J} (k , \tau) = 0 \, .
\label{eq-motion-v}
\end{equation}
Observe that there must exist  $\ntot$ independent solutions $v_{\alpha }^I (k, \tau)$ to this equation (see Appendix~\ref{commutation-app} for a detailed discussion on the $v_{\alpha }^I (k, \tau)$-functions).

Of course, a critical issue here is to set the correct initial conditions for the mode amplitudes $v_{\alpha }^I (k, \tau)$ in such a way that the commutation relations (\ref{commutation-1}) are respected at all times. As a first step towards determining these initial conditions we notice that at a given initial time $\tau = \tau_i$ we may choose each mode $v_{\alpha}^{I} (k , \tau)$  to satisfy the following general initial conditions:
\begin{align}
v_{\alpha}^{I} (k , \tau_i) = & e^I_{\alpha}  v_\alpha(k) \, ,  \label{initial-1}
\\
\frac{ \mathcal D v_{\alpha}^{I} }{dt} (k,\tau_i) = & e^I_{\alpha}  \pi_\alpha(k) \, ,\label{initial-2}
\end{align}
where $e^I_{\alpha}$ is a complete set of unit vectors satisfying $\delta_{I J} e_{\alpha}^I e_{\beta}^J  = \delta_{\alpha \beta}$ and   $\delta^{\alpha \beta} e_{\alpha}^I e_{\beta}^J  = \delta^{I J}$, which should not be confused with the vielbeins defined in (\ref{Q-Q}), and $v_{\alpha}(k)$ and $\pi_{\alpha}(k)$ are factors defining the amplitude of the initial conditions. In order for the commutation relations to be fulfilled, these initial conditions must satisfy:
\begin{equation}
 v_\alpha(k) \pi_{\alpha}^*(k) - v_\alpha^* (k) \pi_{\alpha}(k) = i \, ,
 \label{initial-1-and-2}
\end{equation}
which are the analogous relations to the Wronskian condition in single field slow-roll inflation. Since  the operator ${ \mathcal D }/{d\tau} = d/{d \tau} + Z$ mixes different directions in the $v^I$-field space and since in general the time-dependent matrix $\Omega_{IJ}$ is non-diagonal, then the mode solutions $v_{\alpha }^I (k, \tau)$ satisfying the initial conditions (\ref{initial-1-and-2}) will not remain pointing in the same direction (nor will they remain orthogonal) at an arbitrary time $\tau \neq \tau_i$. In Appendix~\ref{commutation-app} we show that the commutation relations of (\ref{commutation-1}) are consistent with the evolution of the $ v_{\alpha}^{I} (\tau , k)$ dictated by the set of equations of motion (\ref{eq-motion-v}).

In the previous expressions the set of unit vectors $e^I_{\alpha}$ are arbitrary. Moreover, the amplitudes $v_\alpha(k)$ and $\pi_{\alpha}(k)$  entering (\ref{initial-1-and-2}) are in general not uniquely determined, as there is a family of solutions parameterised by the relative phase between $v_\alpha(k)$ and $\pi_{\alpha}(k)$. Indeed, notice that without loss of generality we may write
\begin{equation}
\pi_{\alpha}(k) = \frac{e^{- i \theta_{\alpha}(k)}  }{2  v^*_\alpha(k)  \sin \theta_{\alpha} (k) } \, ,
\end{equation}
where $\theta_{\alpha}(k)$ is a set of real phases relating both amplitudes. Any value for $\theta_{\alpha}(k)$ will satisfy the commutation relations~(\ref{commutation-1}), and therefore they  specify different choices for the vacuum state $| 0 \rangle$. Although in general it is not possible to decide among all the possible values for $\theta_{\alpha}(k)$, fortunately, in the context of inflationary backgrounds $a \rightarrow 0$ as $\tau \rightarrow -\infty$ and a particular choice for these phases becomes handy. Indeed, observe that in the formal limit $a \rightarrow 0$ one has $Z_{IJ} \rightarrow 0$ and $\Omega_{IJ} \rightarrow 0$, which is made explicit by (\ref{def-D-X-2-new}) and (\ref{eom-cov-1}), and the equations of motion (\ref{eq-motion-v}) become:
\begin{equation}
\left( \frac{d^2}{d \tau^2}  + k^2 \right) v_{\alpha}^{I} (k , \tau) = 0 \, .
\label{eq-motion-v-2}
\end{equation}
In this limit there is no mixing between different $\alpha$-modes and perturbations evolve as  if they were in Minkowski background.\footnote{To be more rigorous, in inflationary backgrounds this limit is obtained for $k$-modes such that their wavelength is much smaller than the de Sitter scale $k^2 \gg a^2H^2$.} In this case, we are free to choose $e^{I}_{\alpha} = \delta^I_{\alpha}$ and the solutions to (\ref{eq-motion-v-2}) satisfying the commutation relations  (\ref{commutation-1})  may be chosen as:
\begin{equation}
v_{\alpha}^{I} (k , \tau) = \delta^I_{\alpha} \frac{1}{\sqrt{2 k}} e^{ - i k \tau} \, .
\label{Bunch-Davies}
\end{equation}
Thus we see that in the limit $a \rightarrow 0$ ($\tau \rightarrow - \infty$) we may choose modes in the Bunch-Davies vacuum $\theta_{\alpha} = \pi/2$. We will come back to these conditions in Section~\ref{sec-results} where we set initial conditions on a finite initial time surface
where (\ref{Bunch-Davies}) cannot be exactly imposed.

\subsection{Two-point correlation function}

To finish this general discussion on multi-field perturbations, we proceed to define the spectrum for the perturbations $v^I (\tau,{\bm x})$. The power spectrum, the Fourier transform of the two-point correlation function, is defined in terms of the Fourier modes as
\begin{equation}
\left\langle 0 \left| v^I ({\bm k}, \tau) v^{J*} ({\bm q}, \tau) \right| 0 \right\rangle  \equiv    \delta^{(3)}({\bm k} - {\bm q})  \frac{2 \pi^2}{k^3} \mathcal{P}_v^{I J}  (k ,\tau) \, .
\end{equation}
In terms of the mode amplitudes $v_{\alpha}^{I} (\tau, k)$, this is found to be
\begin{equation}
\mathcal{P}_v^{I J}  (k ,\tau) =   \frac{k^3}{2 \pi^2}  \sum_{\alpha}   v_{\alpha}^{I} (\tau, k)  v_{\alpha}^{J *} (k ,\tau) \, .
\end{equation}
Since the commutation relations require  $\sum_{\alpha} \left[ v_{\alpha}^{I} (k ,\tau)  v_{\alpha}^{J *} (k ,\tau) -  v_{\alpha}^{J} (k ,\tau)  v_{\alpha}^{I *} (k ,\tau) \right] = 0$ (see Appendix~\ref{commutation-app}) we see that the spectrum $\mathcal{P}_v^{I J}  $ is real, as it should be. Additionally, the two point correlation functions in coordinate space may be computed out of $\mathcal{P}_v^{I J}$ as:
\begin{equation}
\left\langle  0 \left| v^I (\tau,{\bm x})  v^J (\tau,{\bm y})  \right| 0 \right\rangle =  \frac{1}{4 \pi} \int \frac{d^3 k}{k^3}   \mathcal{P}_v^{I J}  (k ,\tau) e^{- i {\bm  k} \cdot ( {\bm x} - {\bm y} ) } \, .
\end{equation}
We may also define the power spectrum associated to the $Q^I$ fields instead of the $v^I$ fields. Recalling that $Q^I = v^I /a$, the power spectrum for these fields at a given time $\tau$ is then given by:
\begin{equation}
\mathcal{P}_Q^{I J}  (k ,\tau)  =   \frac{k^3}{2 \pi^2 a^2}  \sum_{\alpha}  v_{\alpha}^{I} (k,\tau)  v_{\alpha}^{I *} (k,\tau) \,  .
\label{power-analytic-1}
\end{equation}
This expression will be used to compute the power spectrum of  the curvature perturbation produced during inflation. Although, in this section we have chosen to exploit a notation whereby Greek indices $\alpha$ label quantum modes, notice that this formalism is equivalent to the use of stochastic Gaussian variables, as in Ref.~\cite{Tsujikawa:2002qx} (see also Ref.~\cite{Lalak:2007vi}).


\section{Models with two scalar fields}
\setcounter{equation}{0}
\label{two-field-power}

\begin{figure}[b!]
\begin{center}
\includegraphics[scale=0.55]{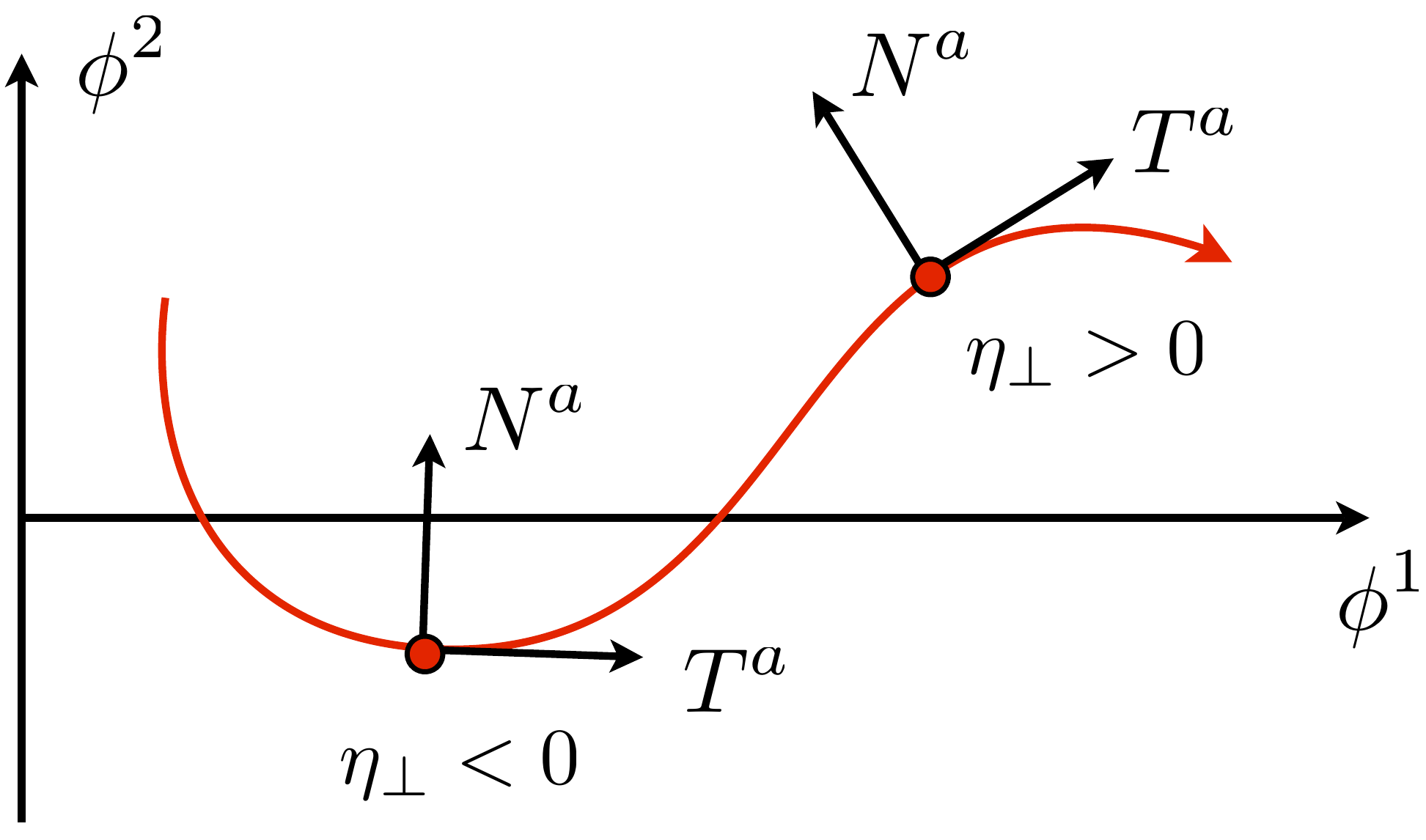}
\caption{\footnotesize The figure shows a fixed right-handed orientation of $N^a$ with respect to $T^a$. If the turn is towards the left then $\eta_{\bot}$ is negative, whereas if the turn is towards the right then $\eta_{\bot}$ is positive.}
\label{fig-N-T-directions}
\end{center}
\end{figure}

We now study the evolution of perturbations in systems containing only two relevant scalar fields. In this case, it is always possible to take the set of vielbeins $\{ e^a_I \}$ to consist entirely in $e^a_T = T^a$ and $e^a_N= N^a$ defined in Section \ref{section-bacground-solution}. Then, the projection tensor $P_{a b}$ introduced in (\ref{D-n})
vanishes identically and one is left with the following relations:
\begin{align}
\frac{D T^a}{d t} = &  - H \eta_{\bot} N^a \, , \label{DT-DN}
\\
\frac{D N^a}{d t} = & \, H \eta_{\bot}  T^a \, .
\label{DN-DT}
\end{align}
At this point we notice that the normal vector $N^a$ has always the same orientation with respect to the curved trajectory, which is due to the presence  of the signature function $s_{N}$ in (\ref{eq:def-t-n}).  For definiteness, let us convene that the normal direction $N^a$ has a  right-handed orientation with respect to $T^a$ as shown in Figure~\ref{fig-N-T-directions}.
With this convention $\eta_{\bot}$  changes signs smoothly in such a way that if the turn is towards the left then $\eta_{\bot}$ is negative, whereas if the turn is towards the right  then $\eta_{\bot}$ is positive. A concrete choice for $T^a$ and $N^a$ with these properties are:
\begin{align}
T^a = & \, \frac{1}{\dot \phi_0} \left( \dot \phi^1 , \dot \phi^2 \right), \label{oriented-T} \\
N^a = & \, \frac{1}{\dot \phi_0 \sqrt{ \gamma} } \left(- \gamma_{22} \dot \phi^2 - \gamma_{12} \dot \phi^1 , \gamma_{11} \dot \phi^1+ \gamma_{21} \dot \phi^2 \right), \label{oriented-N}
\end{align}
where $\gamma = \gamma_{11} \gamma_{22} - \gamma_{12} \gamma_{21}$ is the determinant of $\gamma_{ab}$.
To continue, parallel and normal perturbations with respect to  the inflationary trajectory are then given by
\begin{align}
v^T =  a \, Q^{T} = a \, T_a Q^a \,,
\\
v^N =  a \, Q^{N} = a \, N_a Q^a \,.
\end{align}
By choosing this frame, one finds that $Z_{T N} = - Z_{N T} =  a H \eta_{\bot}$. The coupled equations of motion describing the evolution of both modes $v^T_\alpha (k,\tau)$ and $v^N_\alpha(k,\tau)$ become:
\begin{align}
\frac{d^2 v^T_\alpha}{d \tau^2} + 2 \zeta \frac{ d v^N_\alpha }{d \tau} - \zeta^2  v^T_\alpha +  \frac{d \zeta}{d \tau} v^N_\alpha  + \Omega_{TN} v^N_\alpha + ( \Omega_{TT} + k^2) v^T_\alpha = 0 \,, \label{eq-pert-mode-T}
\\
\frac{d^2 v^N_\alpha}{d \tau^2} - 2 \zeta \frac{ d v^T_\alpha }{d \tau} - \zeta^2  v^N_\alpha - \frac{d \zeta}{d \tau} v^T_\alpha  + \Omega_{NT}v^T_\alpha + (\Omega_{NN} +k^2 ) v^N_\alpha  = 0 \,, \label{eq-pert-mode-N}
\end{align}
where we have defined:
\begin{equation}
\zeta \equiv Z_{T N} =  a H \eta_{\bot} \,.
\end{equation}
In the previous equations, the symmetric matrix $\Omega_{I J}$ is defined in (\ref{definition-C}) and (\ref{eom-cov-1})  and consists of the following elements:
\begin{align}
\Omega_{TT} =&  -  a^2 H^2 \left(  2 + 2 \epsilon -  3  \eta_{|| } + \eta_{||} \xi_{||}  - 4 \epsilon \eta_{||}  + 2  \epsilon^2 -  \eta_{\bot}^2 \right) \,,
\label{omega-TT-final}
\\
\Omega_{NN} =& -  a^2 H^2 (2 - \epsilon) + a^2 M^2 \,,
\label{omega-NN-final}
\\
\Omega_{TN} =&     a^2 H^2 \eta_{\bot} ( 3+ \epsilon - 2  \eta_{||}  -\xi_{\bot}     )  \,,
\label{omega-TN-final}
\end{align}
where $M^2 \equiv V_{N N}  +  H^2 M_{\rm Pl}^{2} \, \epsilon \, \mathbb{R}$ is the effective squared mass of the $v^N$-mode and  $\mathbb{R} = 2 \mathbb{R}_{TNTN} = T^a N^b T^c N^d \mathbb{R}_{a b c d}$ is the Ricci scalar parametrsing the geometry of $\mathcal{M}$. Additionally, we have defined
\begin{equation}
\xi_{\bot} \equiv - \frac{\dot \eta_{\bot}}{H\eta_{\bot}} \,.
\label{def-xi-bot}
\end{equation}
Details on how to arrive at this specific form of $\Omega_{IJ}$ for the case of two-field models are given in Appendix \ref{background-app}.

\subsection{Power spectrum}

Expressions~(\ref{eq-pert-mode-T}) and (\ref{eq-pert-mode-N}) consist of the equations of motion necessary to deduce the generation of  the curvature perturbation
in the case of two-field inflation.
Once the solutions of the fields $v^T = a \, Q^T$ and $v^N = a \, Q^N$ are known, it is possible to define the  curvature
and isocurvature perturbations as
\begin{align}
\mathcal{R} \equiv & \frac{H}{\dot \phi_0} Q^T \, ,
\\
\mathcal{S} \equiv & \frac{H}{\dot \phi_0} Q^N \, ,
\end{align}
respectively. Using equation (\ref{power-analytic-1}) with $I = J = T$, the resulting power spectrum for adiabatic modes are found to be
\begin{equation}
\mathcal{P}_{\mathcal R} (k, \tau) =  \frac{H^2}{\dot \phi_0^2} \mathcal{P}_Q^{TT}  (k ,\tau) = \frac{k^3}{4 \pi^2 a^2 M_{\rm Pl}^2 \epsilon}  \sum_{\alpha=1,2}  v_{\alpha}^{T} (k,\tau)  v_{\alpha}^{T *} (k,\tau)  \, ,
\label{power-analytic-2}
\end{equation}
where $a$ and $\epsilon = \dot \phi^2 / (2 M_{\rm Pl}^2 H^2)$ are functions of $\tau$. We can also compute the power spectrum for isocurvature modes and cross correlation as~\cite{Gordon:2000hv, Amendola:2001ni, Wands:2002bn}
\begin{align}
\mathcal{P}_{\mathcal S} (k, \tau)  = & \frac{H^2}{\dot \phi_0^2} \mathcal{P}_Q^{NN}  (k ,\tau) \, ,
\\
\mathcal{P}_{\mathcal R S} (k, \tau)  = & \frac{H^2}{\dot \phi_0^2} \mathcal{P}_Q^{T N}  (k ,\tau) \, ,
\end{align}
respectively. They can give rise to observable signatures in the CMB power spectrum~\cite{Amendola:2001ni}, but it depends on post-inflationary processes thus we do not consider them here. In this work we are primarily concerned with the computation of the power spectrum of the curvature perturbation
$\mathcal{R}$ at the end of inflation. This corresponds to the quantity
\begin{equation}
\mathcal{P}_{\mathcal R} (k) \equiv \mathcal{P}_{\mathcal R} (k, \tau_{\rm end}) \,, \label{Power-R}
\end{equation}
where $\tau_{\rm end}$ is the time at which inflation effectively
ends\footnote{Since in multi-field inflation the adiabatic mode
  $\mathcal R$ (as well as other background quantities) may continue
  evolving on super horizon scales, here we do not follow the
  standard practice of evaluating the power spectrum at horizon
  crossing time $k = a H$~\cite{Gong:2002cx}. See
  Ref.~\cite{Kinney:2005vj} for a discussion of this point.}. But the
computations of $\calP_\mathcal{S}$ and $\calP_{\calR\mathcal{S}}$ can
be done in an identical way.

\subsection{Effective Theory}
\label{effective-theory-sec}

If a hierarchy of scales is present in the matrix $\Omega_{IJ}$, then we can compute a fairly reliable effective theory out of the system (\ref{eq-pert-mode-T})  and~(\ref{eq-pert-mode-N}). Indeed, by assuming that $\Omega_{NN}$ remains positive at all times and that
\begin{align}
\Omega_{NN} \gg & |\Omega_{TT}| \,,
\\
\Omega_{NN} \gg & |\Omega_{TN}| \, ,
\end{align}
then we may integrate the heavy mode $v^N$ out of the system of equations. By examining the specific shape of the entries $\Omega_{TT}$, $\Omega_{NN}$ and
$\Omega_{TN}$ we see that a generic requisite for this hierarchy to exist is
\begin{equation}
M^2 \gg H^2 \, ,
\end{equation}
where $M^2$ is the effective mass of the heavy mode $v^N$ given by:
\begin{equation}\label{heavymass}
M^2 \equiv V_{N N}  +  H^2 M_{\rm Pl}^{2} \, \epsilon \, \mathbb{R} \, .
\end{equation}
To compute the effective theory we proceed in the same way as in Ref.~\cite{Achucarro:2010jv}. We focus on the mode $\alpha$ associated to slower oscillations due to the hierarchy. Omitting the $\alpha$ label, this mode is necessarily such that
\begin{equation}
\left| \frac{d^2 v^N}{d \tau^2} \right| \ll a^2 M^2 v^N \, .
\end{equation}
This allows us to disregard the second derivative of $v^N$ in (\ref{eq-pert-mode-N}), and write $v^N$ in terms of $v^T$:
\begin{equation}
 v^N  = \frac{1}{\Omega_{NN}  - \zeta^2 + k^2 }\left( 2 \zeta \frac{ d v^T }{d \tau}  + \frac{d \zeta}{d \tau} v^T   - \Omega_{NT}v^T  \right) \, .
 \label{N-ingegration}
\end{equation}
This expression for $v^N$ can be inserted back into the remaining equation of motion~(\ref{eq-pert-mode-T}) to obtain an effective equation of motion for the light adiabatic mode $v^T$. Then, by defining a new field $\varphi$ as
\begin{align}
\varphi \equiv & e^{\beta / 2} v^T \, ,  \label{def-varphi}
\\
e^{\beta(\tau, k^2)} \equiv & 1+ {4 \eta_{\bot}^2} \left( \frac{M^2}{H^2} - 2 + \epsilon - \eta_{\bot}^2 + \frac{k^2}{a^2H^2} \right)^{-1} \,.
\label{betadef1}
\end{align}
we finally arrive at the following effective equation of motion\footnote{Please notice that eqs.~(\ref{effective-theory-eq}) and (\ref{omega-0}) are corrected versions of those appearing in the published version of this article.}
\begin{equation}
\varphi'' + e^{- \beta(\tau,k^2)} k^2 \varphi + \Omega(\tau,k^2) \varphi = 0 \,,  \label{effective-theory-eq}
\end{equation}
where the time dependent  function $\Omega(\tau,k^2)$ is found to be:
\begin{align}
\Omega(\tau,k^2) =& \Omega_0 (\tau) - \frac{\beta''}{2} - \left( \frac{\beta'}{2} \right)^2 - a H \beta' (1 + \epsilon - \eta_{||}) \, , \label{omega-k}
\\
\Omega_0 (\tau)  =&  - a^2 H^2 (2 + 2 \epsilon - 3 \eta_{||} - 4 \epsilon \eta_{||} + \xi_{||} \eta_{||} + 2 \epsilon^2) \, . \label{omega-0}
\end{align}
Notice that $\Omega_0$ is precisely the mass term appearing in the conventional equation of motion for adiabatic fluctuations in single field slow-roll inflation. Furthermore, we note that in the case where the mass $M$ approaches the cutoff of our theory, our results can be derived from an effective action for the adiabatic mode given by the action
\begin{equation}
S = \frac{1}{2} \int d \tau d^3 x \left[  \left( \frac{d \varphi}{d\tau} \right)^2  - \nabla\varphi~ e^{- \beta(\tau, - \nabla^2)}\nabla \varphi - \varphi ~\Omega(\tau,-\nabla^2)\varphi \right] \, , \label{action-effective}
\end{equation}
where $\beta(\tau, - \nabla^2) $ and $\Omega(\tau,- \nabla^2) $ are the functions defined in (\ref{def-varphi}) and (\ref{omega-k}) but with $k^2$ replaced by $-\nabla^2$.
This result corresponds to the generalisation of our previous work~\cite{Achucarro:2010jv} to the case of a slowly rolling background in the presence of gravity. A slightly more formal deduction of this effective theory may be found in appendix \ref{appendix-eft}, where we see that it can be viewed as a leading order effect at the loop level, and as such contains the higher dimensional corrections implied by the general arguments made in Refs.~\cite{Weinberg:2008hq, Cheung:2007st}. In Section~\ref{sec-results} we shall compare the power spectrum obtained using this effective theory with the one obtained from the full set of equations for the perturbations. We anticipate that this effective theory is very reliable regardless of how large the values of $\beta$ are.

\subsection{Slow-roll inflation in two-field models}
\label{section: slow-roll}

So far we have not assumed the slow evolution of background quantities. We now proceed to discuss the case of inflation realised in the slow-roll regime, where the scale of inflation $H$ varies slowly. Our main interest is to study the effects appearing from curved inflationary trajectories, where $\eta_{\bot}$ is non-vanishing. We will assume that the radius of curvature $\kappa$ may take values smaller than $M_{\rm Pl}$, corresponding to turns of the trajectory taking place at field scales smaller than the Planck scale. This situation is certainly allowed and depending on the value of $\epsilon$, it may render large values of $\eta_{\bot}$ (recall (\ref{eta-bot-kappa-2}) relating $\eta_{\bot}$ and $\kappa$). By the same token, we will consider models where the normal mode $v^N$ has a large effective mass $M^2 \gg H^2$.

\subsubsection{Slow-roll parameters}

In general, given the background equations of motion (\ref{scal-eq-1}), (\ref{friedmann-eq-1}) and (\ref{acceleration-eq-1}), we say that a given background quantity $A$ is slowly rolling if its variation satisfies
\begin{equation}
\label{slow-roll-general}
| \delta_{A} | \equiv \left| - \frac{1}{H A} \frac{d A}{d t} \right| \ll 1 \, .
\end{equation}
Observe that  we can write $\epsilon = \delta_{H}$ and $\eta_{||} = \delta_{\dot \phi_0}$, and therefore both $H$ and $\dot \phi_0$ evolve slowly if $\epsilon  \ll 1$ and  $| \eta_{||} | \ll 1$ respectively.  Since $\epsilon = \dot \phi^2 / (2 M_{\rm Pl}^2 H^2)$ then  the condition $|\eta_{||}| \ll 1$ also guaranties that $\epsilon$ will remain varying slowly during inflation. It is useful to introduce  a single small dimensionless number $\delta \ll 1$ parametrising the slow-roll expansion\footnote{Current observations indicate that the order of such a reference parameter is given by the departure of the spectral index from unity $\delta \sim |n_\mathcal{R} - 1|$. We remark here for completeness that the $\epsilon$ and $\eta$ parameters above correspond to the Hamilton-Jacobi slow roll parameters.}, and demand that any quantity $A$ to which slow-roll is imposed, generically satisfies
\begin{equation}
 \frac{1}{H A} \frac{d A}{d t}  =  \mathcal{O} (\delta) \, ,
\end{equation}
which means $\epsilon = \mathcal{O} (\delta)$ and $\eta_{||} = \mathcal{O} (\delta)$. In the absence of clear evidence of it, for simplicity we shall not consider here hierarchies between different slow-roll parameters. Recall that (\ref{first-slow-cond}) and  (\ref{second-slow-cond}) are exact equations relating the parameters $\epsilon$, $\eta_{||}$ and $\xi_{||}$ to the shape of the potential $V$ along the inflationary trajectory. Now, provided that all of these parameters are small, we may re-express these equations to leading order in $\delta$:
\begin{align}
\eta_{||} + \epsilon = & M_{\rm Pl}^2 \frac{\nabla_{\phi} V_{\phi}}{V} \,, \label{eta-pot}
\\
\epsilon = & \frac{M_{\rm Pl}^2}{2} \left( \frac{V_{\phi}}{V} \right)^2 \,. \label{epsilon-pot}
\end{align}
These are the usual equations defining the slow-roll parameters in terms of the shape of the first and second derivatives of $V$.\footnote{Let us recall that  the parameter $\eta$ was originally introduced in the study of single field slow-roll inflation~\cite{Liddle:1992wi}  as $\eta = M_{\mathrm{Pl}}^2 V''/V$. Therefore, in order to compare the present results with those following the original convention, we must write  $\eta= \eta_{||}  + \epsilon$.}  As long as $\epsilon \ll 1$ and $|\eta_{||}| \ll 1$, the background geometry evolves slowly and the scalar field velocity is determined by the attractor equation of motion $3 H \dot \phi_0 +V_{\phi} = 0$.  For completeness, notice from the definition of $\eta_{\bot}$ in (\ref{deff-slow-roll-parameters-3}) that it is possible to write $\eta_{\bot} = V_N / (\sqrt{2 \epsilon} M_{\mathrm{Pl}} H^2)$. Then, using (\ref{epsilon-pot}) we deduce
\begin{equation}
\eta_{\bot}^2 = 9 \left( \frac{V_N}{V_{\phi}}  \right)^2 \, ,
\end{equation}
which is valid to leading order in $\delta$. This equation nicely relates the slope of the potential $V_{\phi}$ along the tangential direction $T^a$ with its counterpart $V_{N}$ along the normal direction $N^a$.

\subsubsection{Perpendicular dynamics}
\label{perpendicular-dynamics}

Let us now turn our attention to parameter $\eta_{\bot}$ defined in (\ref{deff-slow-roll-parameters-3}). Notice that this parameter is {\em not} related to the slow-roll variation of any given background quantity $A$ in the sense of (\ref{slow-roll-general}), and therefore it is not constrained to be of $\mathcal{O}(\delta)$. Moreover, (\ref{eta-bot-kappa-2}) tells us that $\eta_{\bot}$ may be large compared to $\delta$ provided that the radius of curvature $\kappa$ is small compared to $\sqrt{2 \epsilon} M_{\rm Pl}$.
It is important to recognise that the curved inflationary trajectory ($\kappa^{-1} \neq 0$)
has its origin in both the shape of the scalar potential $V$ and the geometry of the scalar manifold where the theory lives. In particular, since $H$ and $\dot \phi_0$ are assumed to evolve slowly, then we expect the flat inflationary trajectory to remain close to the locus of points minimising the heaviest direction $N^a$ of the potential. In other words, to ensure a bending of the trajectory we consider models where the potential is such that
\begin{equation}
V_{NN} \gg | \nabla_{\phi} V_{\phi} | \, .
\end{equation}
It is entirely clear that in the event that the inflationary trajectory is suffering a turn, it will not coincide exactly with curve minimising the heaviest direction, which is made explicit by the result $V_{N} =  \eta_{\bot} \dot \phi_0 H$ found in (\ref{deff-slow-roll-parameters-3}). It is in fact easy to show that the departure $\Delta$ from the real minima $V_N |_{\rm min}= 0$ is roughly given by the condition $V_N + M^2 \Delta \simeq 0$, with $M^2$ given by (\ref{heavymass}). Then, with the help of (\ref{deff-slow-roll-parameters-3}) one finds that the ratio between the deviation $\Delta$ and the radius of curvature $\kappa$ is given by
\begin{equation}
\frac{\Delta }{\kappa} \simeq \eta_{\bot}^2 \frac{H^2}{M^2} \, .
\end{equation}
Observe that $\Delta / \kappa$ is essentially the combination $e^\beta-1$ defined in (\ref{def-varphi}) in the regime $k^2 \ll a^2H^2$. Thus the parameter $\beta$ appearing in the effective theory deduced in Section~\ref{effective-theory-sec} is giving us information regarding the dynamics perpendicular to the inflaton trajectory.

It is important to check out whether the bending interferes with the flatness of the potential as felt by the adiabatic mode $v^T$. Observe from (\ref{eq-pert-mode-T}) and (\ref{omega-TT-final}) that the effective mass $m^2(\tau)$ of $v^T$ is given by
\begin{equation}
m^2(\tau)  \equiv  \Omega_{TT} - \zeta^2 \approx  -  a^2 H^2 (2 + 2 \epsilon - 3 \eta_{||}) \, ,
\end{equation}
where we have neglected terms of $\mathcal{O}(\delta^2)$. Note that $m^2(\tau) = \Omega_0(\tau)$, where $\Omega_0(\tau)$ is the effective mass encountered in the effective theory deduced in Section~\ref{effective-theory-sec}. Thus, we see that $\eta_{\bot}$ does not directly spoil the flatness of the potential $V$. Of course, one should explicitly verify in which way a bending affects the value of $\epsilon$ and $\eta_{||}$ by examining the evolution of the background. We however point out that there is no reason {\it a priori} for which fast and sudden turns with large values of $\eta_{\bot}$ are not possible while staying in the slow-roll regime.

\subsubsection{Equations of motion in the slow-roll regime}

Putting all of the previous results together back into the set of equations ~(\ref{eq-pert-mode-T}) and (\ref{eq-pert-mode-N}), and neglecting terms of $\mathcal{O}(\delta^2)$, we finally arrive at the following equations of motion for the perturbations $v^T_\alpha$ and $v^N_\alpha$:
\begin{align}
\frac{d^2 v^T_\alpha}{d \tau^2} + 2 a H \eta_{\bot} \frac{ d v^N_\alpha }{d \tau}   +   a^2 H^2 \left(  \frac{k^2}{a^2 H^2}  -  2 - 2 \epsilon +  3  \eta_{||} \right) v^T_\alpha  + 2 a^2 H^2  \eta_{\bot}  \left( 2 - \xi_{\bot} \right) v^N_\alpha  =& 0 \,,
\label{eq-pert-mode-T-again}
\\
\frac{d^2 v^N_\alpha}{d \tau^2} - 2 a H \eta_{\bot} \frac{ d v^T_\alpha }{d \tau}  + a^2 H^2  \left(  \frac{k^2}{a^2 H^2} + \frac{M^2}{H^2}  -  2   + \epsilon  - \eta_{\bot}^2 \right) v^N_\alpha +  2 a^2 H^2  \eta_{\bot}    v^T_\alpha  =& 0 \,,
\label{eq-pert-mode-N-again}
\end{align}
where $\xi_{\bot}$ was defined in (\ref{def-xi-bot}). In the next section we deal with these equations numerically for suitable choices of the background parameters, and compare the obtained power spectrum with that of the effective theory obtained in Section~\ref{effective-theory-sec}. We shall see how features in the power spectrum appear as a consequence of curved inflationary trajectory.


\section{Features in the power spectrum}
\setcounter{equation}{0}
\label{sec-results}

We now study the evolution of perturbations and analyse how features in the primordial spectrum are generated  along curved trajectories.
To this extent, we solve (\ref{eq-pert-mode-T-again}) and~(\ref{eq-pert-mode-N-again}) numerically for different background solutions representing curved trajectories and obtain the mode solutions $v_\alpha^I$ which, with the help of (\ref{Power-R}), provide us the desired power spectrum at the end of inflation. For definiteness, we consider models of inflation with an inflationary period of at least $60$ $e$-folds and set the initial  conditions a few $e$-folds before this period starts. To avoid unnecessary complications with initial conditions, we considered models where turns in the trajectory only happen within the last $60$ $e$-folds. Before this period, $\eta_{\bot}=0$ and the equations of motion determining the evolution of perturbations reduce to
\begin{align}
\frac{d^2 v^T_\alpha}{d \tau^2}   +   a^2 H^2 \left(  \frac{k^2}{a^2 H^2}  -  2 - 2 \epsilon +  3  \eta_{||} \right) v^T_\alpha  =& 0 \,,
\label{eq-pert-mode-T-no-turn} \\
\frac{d^2 v^N_\alpha}{d \tau^2}  + a^2 H^2  \left(  \frac{k^2}{a^2 H^2} + \frac{M^2}{H^2}  -  2   + \epsilon  \right) v^N_\alpha   =& 0\,.   \label{eq-pert-mode-N-no-turn}
\end{align}
Then, as long as $\epsilon$ and $\eta_{||}$ are small, we are allowed to make use of initial conditions~(\ref{initial-1}),~(\ref{initial-2}) and~(\ref{initial-1-and-2})  with $e^I_\alpha = \delta^I_\alpha$, and $v_1(k)$ and $v_2(k)$  given by
\begin{align}
v_1(k) = & \frac{\sqrt{\pi}}{4 \sqrt{(1 - \epsilon) a_iH_i}} e^{i \frac{\pi}{2}\left( \nu_1+\frac{1}{2} \right)} H_{\nu_1}^{(1)} \left( \frac{k}{ (1 - \epsilon) a_iH_i} \right) \,,  \\
v_2(k) = & \frac{\sqrt{\pi}}{4 \sqrt{(1 - \epsilon) a_iH_i}} e^{i \frac{\pi}{2}\left( \nu_2+\frac{1}{2} \right)} H_{\nu_2}^{(1)}  \left( \frac{k}{(1 - \epsilon) a_iH_i} \right) \,,
\end{align}
where $H_{\nu}^{(1)}(x)$ denotes the first kind Hankel function, whereas $a_i$ and $H_i$ are the values for the scale factor and Hubble parameter at the initial time $\tau_i$. Similarly, the quantities $\pi_1(k)$ and $\pi_2(k)$ entering the initial conditions~(\ref{initial-1-and-2})  are given by the time derivatives of the previous expressions. On the other hand, the parameters $\nu_1$ and $\nu_2$ are respectively given by:
\begin{align}
\nu_1 & = \sqrt{ \frac{(3 - \epsilon)^2 }{4 (1 - \epsilon)^2} -  \frac{ 3 (\eta - \epsilon)}{(1-\epsilon)^2} } \,,
\\
\nu_2 & = \sqrt{ \frac{(3 - \epsilon)^2 }{4 (1 - \epsilon)^2} - \frac{M^2}{(1-\epsilon)^2  H_i^2} } \,.
\end{align}
Note that in the short wavelength limit $k \gg a_i H_i$, the previous conditions match the mode fluctuations about a Bunch-Davies vacuum (\ref{Bunch-Davies}) discussed in Section~\ref{quant-and-initial}. In all of the cases examined, we consider inflationary trajectories where $\epsilon$, $\eta_{||}$ and $\xi_{||}$ remain small during the interval of interest, while allowing different types of time variation of $\eta_{\bot}$, which is the quantity that parameterises the bending.

\subsection{Constant radius of curvature}

Let us start by considering the simple case in which $\eta_{\bot}$ is
constant during the whole period of inflation where currently
accessible modes were generated. As we have already emphasised, if
$\epsilon$ remains nearly constant, then a constant $\eta_{\bot}$
corresponds to a trajectory with a constant radius of curvature
$\kappa$. We find that the overall effect of having a constant turn is
simply to normalise the amplitude of the spectrum, without modifying
the usual single field dependence of the spectral index $n_\calR$ in
terms of the slow-roll parameters $\epsilon$ and $\eta_{||}$ (see also
\cite{Chen:2009we, Chen:2009zp}):
\begin{equation}
n_\calR  -  1 = 2 \eta_{||} - 4 \epsilon \,.
\end{equation}
In the case $M^2/ H^2 \gg 1$, the predicted power spectrum obtained by the effective theory is indistinguishable from the one obtained by solving the full set of equations. Moreover, with the help of this effective theory, it is in fact possible to infer a simple relation between the power spectrum $\calP_{\mathcal R}(k)$ with $\eta_\bot \neq 0$ and the analytical power spectrum $\calP^{(0)}_{\mathcal R}(k)$ computed with $\eta_\bot = 0$. To this extent, notice that although $\beta(k,\tau)$ is a function of $k$, we see that  after the physical wavelength of the mode becomes larger than the scale $M^{-1}$ (i.e. $k^2/ a^2 \le M^2$), the parameter $\beta(\tau,k)$ becomes effectively $k$ independent, and we can write
\begin{equation}
e^{\beta} = 1 + 4 \eta_{\bot}^2 \frac{ H^2}{M^2} .
\end{equation}
Since $M^2 \gg H^2$, this happens before horizon crossing, and the relevant dynamics is well described by this $k$-independent form of $\beta$. Then the relation between  $\calP_{\mathcal R}(k)$  and $\calP^{(0)}_{\mathcal R}(k)$, as predicted by the effective theory,  becomes:
\begin{equation}
\calP_{\mathcal R}(k) =  \left(1 + 4 \eta_{\bot}^2 \frac{ H^2}{M^2} \right) \calP^{(0)}_{\mathcal R}(k) \,. \label{spectrum-modification}
\end{equation}
This result modifies the usual normalisation condition of the spectrum coming from the COBE data, leading to the following relation among the various parameters:
\begin{equation}
\left(1 + 4 \eta_{\bot}^2 \frac{ H^2}{M^2} \right) \calP^{(0)}_{\mathcal R}(k_\text{COBE}) \approx 2.46 \times 10^{-9} \, .
\end{equation}
Physically, this result may be interpreted  as coming from the fact that heavy and light modes are interchanging energy  at a constant
rate, therefore rendering only a change in the overall amplitude of the spectrum. However, as manifest from the effective theory~(\ref{effective-theory-eq}) the speed of sound is modified as:
\begin{equation}
c_s^2 = e^{-\beta} = \left( 1 + 4 \eta_{\bot}^2 \frac{ H^2}{M^2} \right)^{-1}.
\end{equation}
This implies the generation of non-Gaussianity noticeable in the bispectrum, as studied in Ref.~\cite{Chen:2009we, Chen:2009zp}.

\subsection{Single turn in the trajectory}

\begin{figure}[t!]
\begin{center}
\includegraphics[scale=0.55]{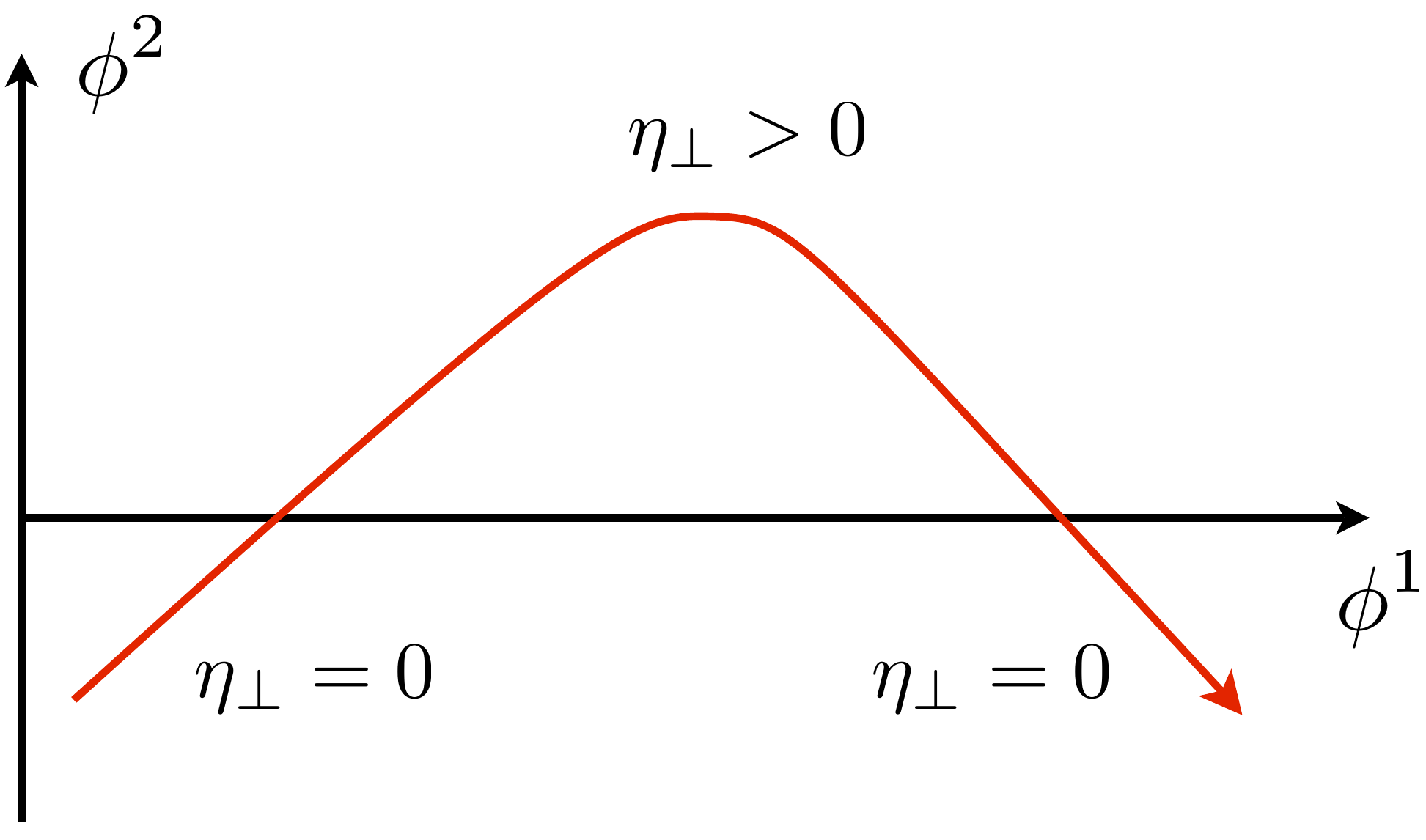}
\caption{\footnotesize The figure shows a prototype example of a trajectory which suffers a localised bend towards the right.}
\label{fig-single-example}
\end{center}
\end{figure}

\begin{figure}[ht!]
\begin{center}
\includegraphics[scale=0.589]{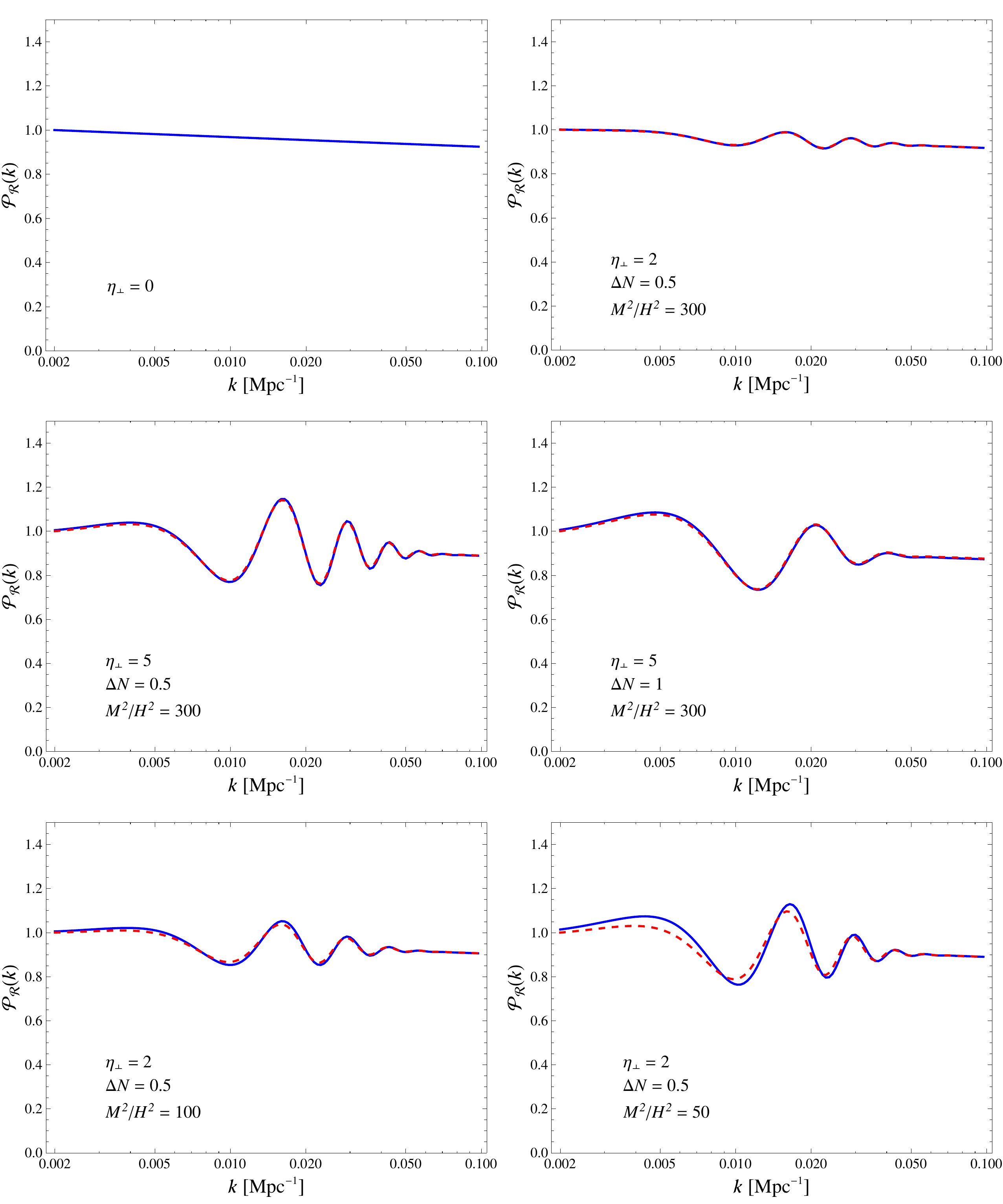}
\caption{\footnotesize The primordial power spectrum $\calP_{\mathcal R}(k)$ normalised in units of $2.46 \times 10^{-9}$, obtained for six different choices of $\Delta N$, $\eta_{\bot {\rm max}}$ and $M^2$. The plots show a comparison between the power spectrum obtained using the full system of equations (solid line) and the one obtained using the effective theory (dashed line). We have chosen as a pivot scale the value $k_{*} = 0.002$Mpc$^{-1}$.}
\label{figura-3}
\end{center}
\end{figure}

As a next step, we consider the presence of a single turn in the inflationary trajectory. To simplify our analysis, we consider the specific case in which the trajectory is initially geodesic (a straight path), then goes through a short period in which it suffers a turn, and finally goes back to a geodesic state. Figure \ref{fig-single-example} shows a prototype  example of such a situation.
We also assume that throughout this process all the slow-roll parameters except for $\eta_\bot$ remain nearly constant.
To model this situation, we take $\eta_{\bot}$ to be an analytical function of the $e$-fold number $N$:
\begin{equation}
\eta_{\bot} (N) = \frac{ \eta_{\bot {\rm max}} }{\cosh^2 \left[ 2 (N - N_0) / \Delta N \right] } \, , \label{eta-bot-N}
\end{equation}
where $\Delta N$ is the number of $e$-folds during which the bending happens, and $N_0$ is the $e$-fold value at which the bending is at its peak, in which case $\eta_{\bot}(N_0) = \eta_{\bot {\rm max}}$. We recall that $N$ may be suitably defined from conformal time $\tau$ through the relation $dN =   a H  d\tau$.
For the other slow-roll parameters, we choose the reference values $\epsilon = 0.022$ and $\eta_{||} = 0.034$. These values correspond to a spectral index $n_\calR = 0.98$, and to a tensor to scalar ratio $r = 0.35$, which are marginally
compatible with current CMB tests~\cite{Larson:2010gs}.
Additionally, these values imply $H = 10^{-5} M_{\rm Pl}$. Figure~\ref{figura-3} shows the power spectra for eight cases with different choices of the parameters $\Delta N$, $\eta_{\bot {\rm max}}$ and $M^2$. The plots\footnote{Please note that the value of $\Delta N$ in these plots corrects a factor of 2 error appearing in the published version.} contain both the spectrum obtained by solving the full coupled system of equations (solid line)
and the spectrum obtained by solving the effective single field equation of motion (dashed line).
For simplicity, we normalise our results in units of $2.46 \times 10^{-9}$
and give the scale $k$  in units of Mpc$^{-1}$. As a reference, we have included the case $\eta_{\bot} = 0$, which corresponds to the power spectrum that would be obtained in the single field case.

The main characteristic shown by the plots are oscillatory features appearing in the spectrum. It may be noticed that the $e$-fold width $\Delta N$ during which the turn takes place actually set the scale $k$ of the oscillatory features. On the other hand, the amplitude of the oscillations is roughly dictated by the ratio $4 \eta_{\bot {\rm max}} H^2 / M^2$. More precisely, the amplitude of the largest oscillatory feature is of order  $\delta \calP_{\mathcal R} / \calP_{\mathcal R} \sim 4 \eta_{\bot {\rm max}} H^2 / M^2$, which agrees with the result of (\ref{spectrum-modification}). Additionally, the match between the curve predicted by the effective theory and the full set of equations becomes better as $M^2/H^2$ acquires larger values, irrespective of how large is $\beta$. In fact, in all of the examples shown we have $\beta \sim 1$.

 The appearance of oscillatory features, not just a single bump, in the spectrum
reflects the fact that both modes $v^N$ and $v^T$ backreact at sub-horizon scales as the turn happens. Once both modes cross the horizon, the amplitude of the adiabatic mode becomes frozen (therefore capturing the moment in which the mode was receiving or releasing energy) while the amplitude of the heavy mode quickly decays due to the accelerated expansion. In fact, we have checked that the levels of isocurvature perturbations at the end of inflation are negligible.

\subsection{A specific example}

\begin{figure}[h!]
\begin{center}
\includegraphics[scale=0.54]{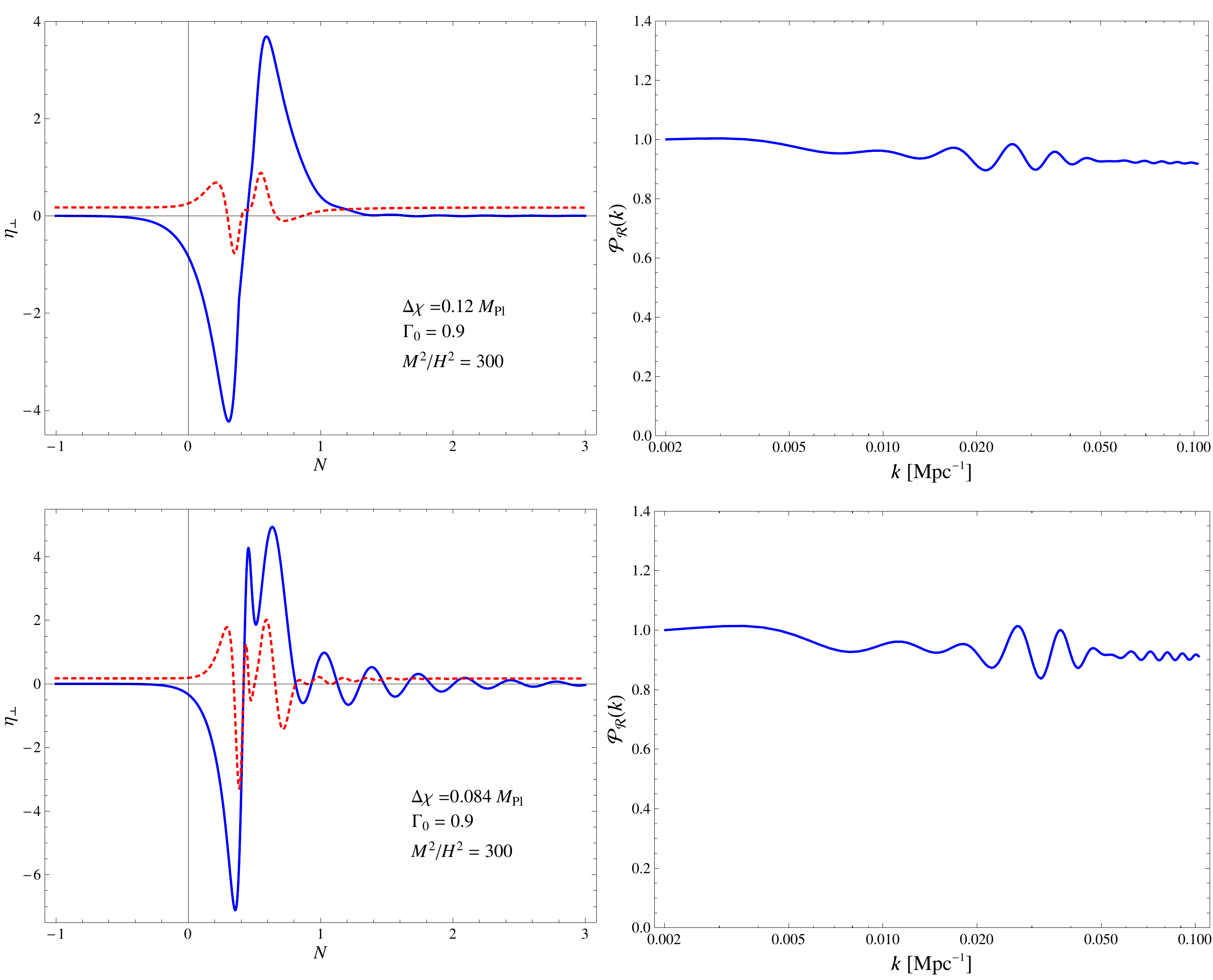}
\caption{\footnotesize Left panels: the evolution of  $\eta_{\bot}$ (solid line) and $5 \times \eta_{||}$ (dashed line) as functions of $e$-fold number $N$  for two set of values of parameters $\Delta \chi$, $\Gamma_0$ and $M^2/H^2$.  In the first case, $\Delta \chi = 0.12 M_{\rm Pl}$ and the maximum value of $\eta_{\bot}$ is about $|\eta_{\bot}| \approx 4$, whereas in the second case, $\Delta \chi = 0.084 M_{\rm Pl}$ and the maximum value becomes $|\eta_{\bot}| \approx 6.5$. Right panels: the resulting primordial power spectrum $\calP_{\mathcal R}(k)$, normalised in units of $2.46 \times 10^{-9}$, obtained for the set of parameters used in the plots of $\eta_{\bot}$. The scale $k$ appears in units of Mpc$^{-1}$.}
\label{fig-plots-toy-model}
\end{center}
\end{figure}

As a last step towards understanding the effects of curved trajectories, we discuss our results applied to a specific toy model, where turns are produced due to the non-trivial evolution of the sigma model metric. Let us consider a two-field model with fields $\phi^1 = \chi$ and $\phi^2 = \psi$ with a kinetic term containing the following sigma model metric:
\begin{equation}
\gamma_{a b} = \left(\begin{array}{cc}1 & \Gamma(\chi) \\ \Gamma(\chi) & 1\end{array}\right) \,,
\end{equation}
where $\Gamma(\chi)$ is only a function of the $\chi$ field and restricted to satisfy $\Gamma^2(\chi) < 1$. The non-vanishing connections are $\Gamma^{\chi}_{\chi \chi} = - \Gamma \Gamma_{\chi}/ (1 - \Gamma^2)$ and $\Gamma^{\psi}_{\chi \chi} = \Gamma_{\chi} / (1 - \Gamma^2)$ with $\Gamma_\chi = \partial_{\chi} \Gamma$, and the equations of motion for the background fields are found to be
\begin{align}
\ddot \chi - \frac{\Gamma \Gamma_{\chi}}{1 - \Gamma^2} \dot \chi^2 + 3H \dot \chi + \frac{1}{1-\Gamma^2} V_{\chi} - \frac{\Gamma}{1-\Gamma^2} V_{\psi} =& 0 \,, \\
\ddot \psi + \frac{ \Gamma_{\chi}}{1 - \Gamma^2} \dot \chi^2 + 3H \dot \psi + \frac{1}{1-\Gamma^2} V_{\psi} - \frac{\Gamma}{1-\Gamma^2} V_{\chi} =& 0 \,,
\end{align}
 where $V_\chi = \partial_{\chi} V$ and $V_\psi = \partial_{\psi} V$.
For concreteness, let us consider the following separable scalar field potential:
\begin{equation}
V(\chi,\psi) = V_0 (\chi)+ \frac{1}{2} M^2 \psi^2 \,.
\end{equation}
In the particular case of $\Gamma = 0$, the dynamics of the two fields decouple and inflation may be achieved with $\chi$ by a suitable choice of the potential $V_0 (\chi)$. If, however, $\Gamma(\chi)$ is allowed to be non-vanishing for certain values of  $\chi$, then a mixing between the two modes is inevitable, and the inflationary trajectory will  be curved. Following the discussion at the beginning of Section~\ref{two-field-power}, we choose the tangential and normal vectors $T^a$ and $N^a$ as in  (\ref{oriented-T}) and (\ref{oriented-N}):
\begin{align}
T^a = & \, \frac{1}{\dot \phi_0} \left ( \dot \chi ,  \dot \psi \right), \\
N^a = & \, \frac{1}{\dot \phi_0 \sqrt{1 - \Gamma^2} } \left(-\dot \psi - \Gamma \dot \chi, \dot \chi + \Gamma \dot \psi \right),
\end{align}
where $\dot \phi_0 = \dot \chi^2 + \dot \psi^2 + 2 \Gamma \dot \chi \dot \psi$. Recall that with this convention $\eta_{\bot}$ is allowed to change its sign. The relevant background parameters describing this situation are then:
\begin{align}
\epsilon = & \, \frac{ \dot \chi^2 + \dot \psi^2 + 2 \Gamma \dot \chi \dot \psi}{2 M^2_{\rm Pl} H^2} \,,
\\
\eta_{||} = & \, 3 + \frac{\dot \chi V_\chi + \dot \psi V_\psi}{H \left(\dot \chi^2 + \dot \psi^2 + 2 \Gamma \dot \chi \dot \psi\right)} \,,
\\
\eta_{\bot}  = & \, - \frac{\left(\dot \psi + \Gamma \dot \chi\right) V_\chi - \left(\dot \chi + \Gamma \dot \psi\right) V_\psi}{H \sqrt{1-\Gamma^2} \left(\dot \chi^2 + \dot \psi^2 + 2 \Gamma \dot \chi \dot \psi\right)} \,,
\end{align}
where $H$ is given by $6 M^2_{\rm Pl} H^2 = \dot \chi^2 + \dot \psi^2 + 2 \Gamma \dot \chi \dot \psi + 2 V $.
For concreteness, let us consider a parameter $\Gamma(\chi)$ having the following $\chi$-dependence
\begin{equation}
\Gamma(\chi) = \frac{\Gamma_{0}}{\cosh^2 \left[ 2 (\chi - \chi_0)/\Delta\chi \right] } \,,
\end{equation}
where $\Gamma_0$ is the maximum value attained by $\Gamma(\chi)$. On the other hand, we take a generic potential $V_0(\chi)$ rendering values $\epsilon = 0.022$ and $\eta_{||} = 0.034$ for the slow-roll parameters in the absence of curves. For this specific configuration, we found that the background value of $\epsilon (\tau)$ remains nearly constant at the attractor value  $\epsilon = 0.022$  whereas the background value of $\eta_{||}(\tau)$ is more sensitive to the turns suffered by the trajectory, having small deviations from the attractor value $\eta_{||} = 0.034$. Additionally, we found two relevant time scales determining the behaviour of background quantities   $\eta_{||}$ and $\eta_{\bot}$:
\begin{align}
T_{\psi} \equiv & M^{-1} \,,
\\
T_{\chi}  \equiv & \frac{\Delta \chi }{ \dot \phi_0 }=  \frac{ \Delta \chi }{\sqrt{2 \epsilon} M_{\rm Pl} H }\,.
\end{align}
The appearance of these time scales are actually easy to understand. First, notice that $T_{\chi}$ is the time during which the turn takes place whereas $T_{\psi}$ is the oscillation period of the massive field $\psi$. We find that if  $T_{\chi} \ll T_{\psi}$, then the background dynamics is such that $\phi^a_0 = (\chi, \psi)$ oscillates about $\psi = 0$, meaning that both $\eta_{||}$ and $\eta_{\bot}$ presented oscillatory features with frequency $\mathcal{O}\left(T^{-1}_{\psi}\right)$. On the other hand, if $T_{\chi} \gg T_{\psi}$, the background field departs adiabatically from the minima of the potential $\psi = 0$, and  the time evolution of $\eta_{||}$ and $\eta_{\bot}$ is dictated by the time scale $T_{\chi}$. This latter case may be interpreted as a situation where the trajectory is momentarily pushed towards one of the walls of the potential, as the curve takes place. Figure~\ref{fig-plots-toy-model} shows the background values of  $\eta_{\bot}$ and $\eta_{||}$ (as functions of the $e$-fold number $N$) for the case $\Gamma_0 = 0.9$, $M^2 = 300 H^2$ and two values of $\Delta \chi$, namely $\Delta \chi = 0.12 M_{\rm Pl}$ and $\Delta \chi = 0.084 M_{\rm Pl}$.\footnote{Notice that these plots are corrected versions of those appearing in the published version of this article.} In the latter case, it may be appreciated how the time scale $T_{\psi}$ appears in the shape of $\eta_{\bot}$. 

Figure~\ref{fig-plots-toy-model} also shows the power spectrum obtained for the two described cases (right panels). In the present examples, the features appearing in the spectrum are not as regular as those of Figure~\ref{figura-3}. This is mainly because in the present situation the curvilinear trajectory contains several turns, in order to go back to the attractor solution.  Although in this specific model the slow-roll parameter $\eta_{||}$ appears to be sensitive to the mass scale $M$ and the curves taking place, it is important to notice that this is a model dependent characteristic, and that in general $\eta_{||}$ may show various types of behaviour depending on the sigma model metric and the potential.  In general, however, the momentary time variation of $\eta_{||}$ due to curved trajectories does not spoil the slow-roll regime, and background fields tend to quickly evolve back to the attractor behaviour characteristic of the single field case as soon as the bending of the trajectory stops. In this regard, we find that the time variation of $\eta_{||}$ is not relevant for the appearance of features in the power spectrum, and that the main contribution is that coming from the derivative interactions due to $\eta_{\bot}$ in the equations of motion. 

\subsection{Enhancement of non-Gaussianity}

We briefly elaborate here on another potentially observable feature not already discussed. The previous section computed the power spectrum of the curvature perturbation for a few examples where the inflaton traverses sufficiently curved regions in field space. From the results, it is clear that features in the spectrum will result each time the trajectory traverses a bend. These features are produced via the kinetic interaction between the heavy isocurvature modes and the light curvature mode as the turns are traversed by the background field. Crucially in these examples the heavy mode remained very massive throughout ($M^2 \gg H^2$), highlighting the fact that heavy fields may not always be disregarded (truncated) when computing the spectrum for adiabatic modes.

What is important to note is that the interaction between curvature and isocurvature modes implies a change in the speed of sound for the curvature perturbations-- as long as $M^2 \gg H^2$, then $\beta(\tau,k)$ is effectively $k$-independent before horizon crossing and the speed of sound may be written as
\begin{equation}
c_s^2 = e^{-\beta} = \left( 1 + 4 \eta_{\bot}^2 \frac{H^2}{M^2} \right)^{-1} \, .
\end{equation}
As is well known, a model with a speed of sound significantly smaller than unity gives rise to a noticeable level of non-Gaussianity of equilateral type, characterised by the non-linear parameter~\cite{Bartolo:2004if}:
\begin{equation}\label{equi_fNL}
f_\mathrm{NL}^\text{(eq)} \sim \frac{1}{c_s^2} \, .
\end{equation}
Thus, we are led to reason that for generic models of inflation with curvilinear trajectories in a multi-dimensional field space, glitches in the power spectrum are accompanied by a correlated enhancement of non-Gaussianity of the equilateral type, provided that the turns in the inflaton trajectory violate the adiabatic approximation vigorously enough-- a phenomenon which we have argued occurs at various points in field space in many realistic realisations of inflation. Thus although there appear to be many models where either non-trivial modulations in the power spectrum (e.g. features in the single field inflaton potential~\cite{Starobinsky:1992ts, Adams:2001vc, TocchiniValentini:2004ht, Gong:2005jr, Covi:2006ci, Hunt:2007dn, Ichiki:2009xs, Peiris:2009wp, Hamann:2009bz}) or large equilateral non-Gaussianity (e.g. DBI inflation~\cite{Silverstein:2003hf, Alishahiha:2004eh}) result, it appears that in generic multi-field models with curved inflationary trajectories, both are present and correlated. Evidently, the effective quadratic action (\ref{action-effective}) contains the leading higher order corrections which can also result in non-Gaussian signatures and implies the non-linear parameter (\ref{equi_fNL})~\cite{Cheung:2007st}. However to fully describe the bispectrum associated with the curvature perturbation, we need to properly take into account the cubic order action including gravity. We will discuss this issue in a separate publication.

\section{Conclusions}
\label{sec: conclusions}

Multi-field models of inflation contain a range of physics which goes
beyond that encountered within the single-field paradigm. In this
work we have focused on the particular case where all of the scalar
fields remain massive during inflation except for one, which slowly
rolls down the multi-field potential. We have found that curved inflationary trajectories can generate significant features in the
primordial spectrum of density perturbations arising from normal modes
becoming excited and backreacting on the dynamics of the adiabatic
mode.

To achieve these results we analysed the evolution of the quantum
perturbations of a general multi-field setup, including the presence
of a non-canonical kinetic term. Our methods are completely general
and naturally incorporate those implemented in previous
works~\cite{Lalak:2007vi, Tsujikawa:2002qx}, where stochastic Gaussian
variables are used. Moreover, although the main focus of this work was
the study of systems where there exists a hierarchy, our results may be
used to study a wide range of situations, including where no such hierarchies
are present.

Our formalism allows us to consider time-dependent situations beyond
the regime of applicability of existing methods, such as inflaton
trajectories with fast, sudden turns (regardless of whether the sigma
model metric is canonical or non-canonical) as well as more general
situations in which the masses of the heavy fields in the orthogonal
direction are changing along the trajectory (even if they still remain
much heavier than $H^2$ and all other scales of interest). Additionally, we wish to emphasise that these non-decoupling effects have their origin in the non-geodesic nature of the trajectories in field space \footnote{The recent reference \cite{Cremonini-arxiv} discusses some of these effects in a particular model, the so-called gelaton model of Ref.~\cite{Tolley:2009fg}. We want to point out that the non-decoupling and the reduced speed of sound discussed in these two papers is not so much due to the non-trivial sigma model metric but to the non-geodesic nature of the trajectories considered.}.

Our results highlight the limitations of simply truncating heavy
physics when modelling single-field realisations of inflation and show
under which circumstances high energy effects can leave an imprint on the power
spectrum. The main reason behind these effects is the existence of
kinetic couplings between adiabatic and non-adiabatic modes, emerging
as the inflationary trajectory suffers a turn. As we have seen in
Section \ref{canonical-frame}, it is always possible to change basis
to a canonical frame where such interactions are absent. In that case,
the eigenvectors of the perturbation mass matrix quickly vary as the
inflationary trajectory turns, and we are left with the alternative
point of view by which these high energy effects appear due to a
violation of the adiabatic condition for truncating heavy fields. In
fact, if the heavy fields are sufficiently massive, we find that we
can construct an effective field theory for the adiabatic modes
encapsulating the relevant effects of the full multi-field
dynamics. As we have seen, such effects are not mere corrections to
the standard single field theory, but represent entirely new
contributions to the quadratic action for perturbations.

We find particularly noteworthy, the presence of potentially
observable signatures that result from a reduced speed of sound for
the adiabatic perturbations during sudden turns. As a corollary,
correlated non-Gaussianity will also manifest as a result of these
sudden turns although a full analysis studying the details of their
appearance in multi-field inflation is beyond the scope of this work
and will be addressed in a future report~\cite{us-non-gaussian}.
Nevertheless, it would appear that in generic multi-field models with
curved inflationary trajectories, both effects are present and
correlated, and can potentially give information about other, much
heavier, fields that would otherwise be inaccessible to experiment.

\section*{Acknowledgements}

We would like to thank Jaume Garriga, Jan Hamann, Simeon Hellerman,
Koenraad Schalm and the participants of the Focus Week on String
Cosmology at IPMU for discussion and comments. Additionally, we thank Sebasti\'an C\'espedes 
for helping us in the correction of Figures 5 and 6 and a few typographical errors. This work was
partially supported by the Netherlands Organisation for Scientific
Research (N.W.O.) under the Vici and Vidi programmes (AA,SH,JG), by
Conicyt under the Fondecyt initiation on research project 11090279
(GAP), by funds from CEFIPRA/IFCPAR project 4104-2 (SP) and by the
Consolider-ingenio programme CDS2007-00042 (AA). GAP wishes to thank
LPTENS, CPHT at the Ecole Polytechnique and the Lorentz Institute
(Leiden) for their hospitality during the preparation of the
manuscript. SP wishes to thank the theory group at Leiden University
for hospitality during the preparation of the manuscript.

\begin{appendix}


\renewcommand{\theequation}{\Alph{section}.\arabic{equation}}
\setcounter{section}{0}
\setcounter{equation}{0}

\section{Additional details on some background quantities}
\label{background-app}

To obtain (\ref{D-n}) we proceed as follows: first, by taking a total time derivative to (\ref{scal-eq-1}) we obtain
\begin{equation}
\frac{1}{\dot \phi_0} \frac{D^2 \dot \phi_0^a}{dt^2} = 3H^2 (\epsilon T^a + \eta^a) -  \nabla_{\phi} V^a \,,
\label{ddot-dotphi}
\end{equation}
where $\nabla_{\phi} \equiv T^a \nabla_{a}$. Recalling that $T^a = \dot \phi^a_0 / \dot  \phi_0$, the previous equation can be re-expressed as
\begin{equation}
 \frac{D^2 T^a}{dt^2} = T^a  \nabla_{\phi} V_{\phi} - \nabla_{\phi} V^{a} - \frac{\left(V_{\phi} - \ddot \phi_0\right) V_N}{\dot \phi_0^2} N^a \,.
\end{equation}
On the other hand, taking a total time derivative to (\ref{eq-mot-t-n-1}) we may obtain yet another expression for the second variation $ {D^2 T^a}/{dt^2} $, given by
\begin{equation}
 \frac{D^2 T^a}{dt^2} =  \left( \frac{ V_N \ddot \phi_0}{\dot \phi_0^2} - \frac{\dot V_N}{\dot \phi_0} \right) N^a  - \frac{V_N}{\dot \phi_0} \frac{D N^a}{dt} \, .
\end{equation}
Equating the last two expressions and performing some straightforward algebraic manipulations, we finally obtain
\begin{equation}
 \frac{D  N^a}{dt} = H \eta_{\bot} T^a + \frac{1}{H \eta_{\bot}}  P^{a b} \nabla_{\phi} V_b \, ,  \label{D-n-app}
\end{equation}
where we have defined the projector tensor $P^{ab} \equiv \gamma^{a b} - T^a T^b - N^a N^b$ along the space orthogonal to the subspace spanned by the unit vectors $T^a$ and $N^a$. That is, $P_{a b} N^b = 0$ and $P_{a b} T^b = 0$.

To arrive at the form of the mass matrix $\Omega_{IJ}$ shown in (\ref{omega-TT-final}), (\ref{omega-NN-final}) and (\ref{omega-TN-final}), we may start from the explicit form deduced out of (\ref{definition-C}) and (\ref{eom-cov-1}) for the case of two-field models:
\begin{align}
\Omega_{TT} =&  -  a^2 H^2 (2 - \epsilon) +  a^2 V_{\phi \phi} - 2 a^2 H^2 \epsilon \left(3 - 2 \eta_{||}  + \epsilon  \right) \,,
\label{omega-TT}
\\
\Omega_{NN} =& -  a^2 H^2 (2 - \epsilon) + a^2 V_{N N}  +  a^2  H^2 M_{\rm Pl}^{2} \, \epsilon \, \mathbb{R} \,,
\label{omega-NN}
\\
\Omega_{TN} =&    a^2 V_{\phi N} + 2 a^2 H^2 \eta_{\bot} \epsilon \,,
\label{omega-TN}
\end{align}
where we have defined
\begin{align}
V_{\phi \phi}  \equiv & T^a T^b \nabla_{a} V_{b} \,,
\\
V_{NN} \equiv & N^a N^b \nabla_a V_{b} \, ,
\\
V_{\phi N} \equiv & T^a N^b \nabla_{a} V_{b} \, .
\end{align}
Additionally $\mathbb{R} = 2 \mathbb{R}_{TNTN} = T^a N^b T^c N^d \mathbb{R}_{a b c d}$ is the Ricci scalar parametrising the geometry of $\mathcal{M}$. Notice that  $V_{\phi \phi}$ can be rewritten in the following way:
\begin{align}
V_{\phi \phi} =& T^a \nabla_a (T^b V_b) - T^a (\nabla_a T^b) V_b  \nonumber\\
=& \nabla_{\phi} V_{\phi} - \frac{1}{\dot \phi_0} \frac{DT^b}{dt} V_b  \nonumber\\
=&  \nabla_{\phi} V_{\phi} + H^2 \eta_{\bot}^2 \,,
\label{TT-V}
\end{align}
where, to go from the second to the third line we made use of (\ref{DT-DN}) and relation $V_N = \dot \phi_0 H \eta_{\bot}$ coming from the definition of $\eta_{\bot}$ in (\ref{deff-slow-roll-parameters-3}). Similarly, the quantity $V_{\phi N}$ may be manipulated in the following way:
\begin{align}
V_{\phi N} =& T^a \nabla_a (N^b V_b) - T^a (\nabla_a N^b) V_b  \nonumber \\
=& \nabla_{\phi} V_N - \frac{1}{\dot \phi_0} \frac{D N^b}{dt} V_b  \nonumber \\
=& \nabla_{\phi} V_N - \frac{ H \eta_{\bot} }{\dot \phi_0} V_\phi \,, \label{TN-V}
\end{align}
where again, to go from the second to the third line, we made use of (\ref{DT-DN}). As a final step, we may use $V_{N} = \dot \phi H \eta_{\bot} $ to deduce:
\begin{equation}
\nabla_{\phi} V_N = \frac{1}{\dot \phi_0} \frac{d}{dt} \left( \dot \phi_0 H \eta_{\bot} \right) = - H^2 \eta_{\bot} (\eta_{||} + \epsilon + \xi_{\bot}) .
\end{equation}
Collecting all of these terms back into  (\ref{omega-TT}), (\ref{omega-NN}) and (\ref{omega-TN}) we finally arrive at  (\ref{omega-TT-final}), (\ref{omega-NN-final}) and (\ref{omega-TN-final}).
Observe that we are not able to rewrite $V_{NN} = N^a N^b \nabla_a V_{b}$ in a similar way, since it involves second variations away from the inflationary trajectory. This simply means that the quantity $V_{NN}$ must be regarded as an additional parameter of the model related to the mass of the transverse mode with respect to the inflaton trajectory.


\section{Commutation relations for quantum multi-fields}
\setcounter{equation}{0}
\label{commutation-app}

In this appendix we show that the commutation relations in (\ref{commutation-1}) are fully consistent with the evolution of the $ v_{\alpha}^{I} (k,\tau)$ dictated by the set of equations of motion~(\ref{eq-motion-v}). To begin with, observe that in order to satisfy the commutation relation~(\ref{commutation-1}) the 
 $\ntot$
mode solutions $v_{ \alpha }^I (k,\tau)$ must satisfy the following conditions:
\begin{align}
\sum_{\alpha}  \left[ v_{\alpha}^I  \frac{ \mathcal D v_{\alpha}^{J *}}{d\tau}   -  \frac{ \mathcal D v_{\alpha}^{J}}{d \tau}  v_{\alpha}^{I *}  \right] &= i \delta^{I J} \,, \label{cond-v-1} \\
\sum_{\alpha}  \left[ v_{\alpha}^I  v_{\alpha}^{J *} -  v_{\alpha}^J  v_{\alpha}^{I *}  \right] &= 0 \,, \label{cond-v-2} \\
\sum_{\alpha}  \left[ \frac{ \mathcal D v_{\alpha}^{I}}{d\tau}  \frac{ \mathcal D v_{\alpha}^{J *}}{d\tau}   - \frac{ \mathcal D v_{\alpha}^{J}}{d\tau}  \frac{ \mathcal D v_{\alpha}^{I *}}{d\tau}  \right] &= 0 \label{cond-v-3} \,.
\end{align}
To show that these relations are satisfied at any given time $t$ we proceed as follows: first, let us define the following matrices:
\begin{align}
A^{I J} &= i \sum_{\alpha} \left[ v_{\alpha}^I  v_{\alpha}^{J *} -  v_{\alpha}^J  v_{\alpha}^{I *}  \right]  \,, \\
B^{I J} &= i \sum_{\alpha} \left[ \frac{ \mathcal D v_{\alpha}^{I}}{d\tau}  \frac{ \mathcal D v_{\alpha}^{J *}}{d\tau}   - \frac{ \mathcal D v_{\alpha}^{J}}{d\tau}  \frac{ \mathcal D v_{\alpha}^{I *}}{d\tau}  \right]  \, , \\
E^{I J} &= i \sum_{\alpha}  \left[ v_{\alpha}^I  \frac{ \mathcal D v_{\alpha}^{J *}}{d\tau}   -  \frac{ \mathcal D v_{\alpha}^{J}}{d\tau}  v_{\alpha}^{I *}  \right]  \,.
\end{align}
These tensors satisfy the following properties
\begin{align}
A^{IJ} = & A^{IJ *} = - A^{JI} \,, \\
B^{IJ} = & B^{IJ*} = - B^{JI}\,, \\
E^{IJ} = & E^{I J*} \,.
\end{align}
In other words, they are real, with $A^{IJ}$ and $B^{IJ}$ antisymmetric ($E^{IJ}$ has no specific symmetries). Because of these properties $A^{IJ}$ and $B^{IJ}$ consist of
 $\ntot(\ntot-1)/2$ independent real components each, whereas $E^{IJ}$ consists of  $\ntot^2$
independent real components. Thus, in order to fix the values of all of these tensors we need to specify  $2\ntot^2-\ntot$
independent quantities. These tensors also satisfy the following equations of motion:
\begin{align}
 \frac{ \mathcal D}{d\tau} A^{IJ} =& E^{IJ} - E^{JI} \,, \label{tensor-1} \\
 \frac{ \mathcal D}{d\tau}  B^{IJ} =& \Omega^{I}{}_{K}E^{KJ} -  \Omega^{J}{}_{K}E^{K I} \,, \label{tensor-2} \\
 \frac{ \mathcal D}{d\tau} E^{IJ} =& B^{IJ} + A^{IK} \left( k^2 \delta^J_K + \Omega_{K}{}^{J} \right) \,. \label{tensor-3}
\end{align}
Taking the trace to the last equation, we obtain that the trace $E \equiv E^{I}{}_{I}$ satisfies
\be
\frac{d E}{d \tau} = 0 \,,
\ee
and therefore $E$ is a constant of motion of the system. Furthermore, observe that the configuration $E^{IJ} = E \delta^{IJ} / 
 \ntot$
and $A^{IJ} = B^{IJ} = 0$ for which conditions (\ref{cond-v-1}) to (\ref{cond-v-3}) are satisfied corresponds to a fixed point of the set of equations (\ref{tensor-1}) to (\ref{tensor-3}). That is, they automatically satisfy:
\be
\frac{\mathcal D}{d\tau} A^{IJ} =\frac{\mathcal D}{d\tau} B^{IJ} = \frac{\mathcal D}{d\tau} E^{IJ} = 0\,.
\ee
Therefore, it only remains to verify whether there exist sufficient independent degrees of freedom in order to satisfy the initial conditions $E^{IJ} = E \delta^{IJ} / \ntot$
and $A^{IJ} = B^{IJ} = 0$ at a given initial time $\tau_{i}$. As a matter of fact, we have exactly the right number of degrees of freedom. As we have already noticed there exists
 $\ntot$ independent solutions $v_{\alpha}^{I} (k,\tau)$ to the equations of motion. To fix each solution $v_{\alpha}^{I} (k,\tau)$ we therefore need to specify
 $2\ntot^2$ independent quantities, corresponding to the addition of  $\ntot^2$
components $v_{\alpha}^{I} (\tau_i)$ and  $\ntot^2$
momenta ${ \mathcal{D} v_{\alpha }^{I}}/{d\tau} (\tau_i)$. However we must notice that the overall phase of each solution $v_{\alpha}^{I} (k,\tau)$ plays no roll in setting the initial values for $A^{IJ}$, $B^{IJ}$ and $E^{IJ}$. We therefore have precisely   $2\ntot^2-\ntot$
free parameters to set $E^{IJ} = E \delta^{IJ} /  \ntot$
and $A^{IJ} = B^{IJ} = 0$. Of course, the value of the trace $E$ is part of this freedom, and we are free to fix it in such a way that $E/\ntot=1$.

To summarise, it is always possible to choose the initial conditions for $v_{\alpha}^{I} ( k,\tau)$ and ${ \mathcal{D} v_{\alpha }^{I}}/{d\tau} (k,\tau)$ in such a way that conditions (\ref{cond-v-1}) to (\ref{cond-v-3}) are satisfied. These conditions ensure the commutation relation~(\ref{commutation-1}). To finish this discussion, recall that one possible choice for the initial conditions for the perturbations  allowing (\ref{cond-v-1}) to (\ref{cond-v-2}) to be satisfied, are precisely those expressed in (\ref{initial-1-and-2}) with suitable choices for the coefficients $ v_\alpha(k)$ and $\pi_{\alpha}(k)$:
\begin{equation}
 v_\alpha(k) \pi_{\alpha}^*(k) - v_\alpha^* (k) \pi_{\alpha}(k) = i  \,,
\end{equation}
 for $\alpha=1,\cdots\ntot$. We should emphasise however that this is not the unique choice for initial conditions, and in general, any choice for which $E^{IJ} = E \delta^{IJ} /  \ntot$ and $A^{IJ} = B^{IJ} = 0$ will do just fine.


\section{Effective theory for the adiabatic mode}
\setcounter{equation}{0}
\label{appendix-eft}

In this appendix we offer another deduction of the effective theory shown in Section \ref{effective-theory-sec}. We begin by writing the action (\ref{action-for-v-fields}) for the particular case of two fields:
\begin{align}
\label{eea}
S =& \int~d\tau d^3x~\frac{1}{2} \left[ \left(\frac{d v^T}{d\tau}\right)^2 - \left(\nabla v^T\right)^2 - \left( \Omega_{TT} - \zeta^2 \right) \left(v^T\right)^2 \right]  \nonumber \\
& + \int~d\tau d^3x~\frac{1}{2}\left[ \left(\frac{d v^N}{d\tau}\right)^2 - \left(\nabla v^N\right)^2 -  \left( \Omega_{NN} - \zeta^2 \right) \left(v^N\right)^2 \right] \nonumber \\
& - \int~d\tau d^3x~v^N \left( \Omega_{T N}  - \frac{d \zeta}{d \tau}   - 2 \zeta \frac{d}{d\tau} \right)  v^T \, .
\end{align}
Given that $\Omega_{NN} \gg  |\Omega_{TT}|$ and $\Omega_{NN} \gg  |\Omega_{TN}|$ the field $v^N$ is the heavier of the two. Taking this scale as the scale of the heavy physics that we wish to integrate out, we can formally evaluate the functional integral for $v^N$ to obtain the one loop effective action for $v^T$ as
\begin{align}
\label{effa}
S =& \int~d\tau d^3x~\frac{1}{2}\left[ \left(\frac{d v^T}{d\tau}\right)^2 - \left(\nabla v^T\right)^2 - ( \Omega_{TT} - \zeta^2) \left(v^T\right)^2 \right]
\nonumber\\
& + \frac{1}{2}\int~d\tau d^3x \int~d\tau'd^3x'\mathcal O(\tau)v^T({ \bm x},\tau)G({ \bm x},\tau;{ \bm x}',\tau')\mathcal O(\tau')v^T({ \bm x}',\tau') ~+~S_{CT} \, ,
\end{align}
with $\mathcal O$ given by
\begin{equation}
\mathcal O(\tau) \equiv -  \left( \Omega_{T N}  - \frac{d \zeta}{d \tau}   - 2 \zeta \frac{d}{d\tau} \right)  \, ,
\end{equation}
and
\begin{equation}
\label{gfm}
G\left(x,\tau;x',\tau'\right) = \frac{1}{\square + \Omega_{NN} - \zeta^2} \, .
\end{equation}
The term $S_{CT}$ renormalises the effective action for the background inflaton field, and we have to demand that the parameters of this effective action that satisfy the slow roll conditions rather than those of the bare action~\cite{Burgess:2010dd}, which we presume to be the case here. In general, evaluating the full effective action is a highly non-trivial task. However in Fourier space, one can formally make the expansion:
\begin{equation}
G\left(\tau,\tau',k \right) = \frac{1}{-\partial_\tau^2 + k^2 + \Omega_{NN} - \zeta^2} = \frac{1}{\omega^2}\Bigl(1 - \frac{\partial_\tau^2}{\omega^2}+\cdots\Bigr) \, ,
\end{equation}
where
\begin{equation}
\omega^2 \equiv k^2 + \Omega_{NN} - \zeta^2.
\end{equation}
Where implicit in the above is that if the scale $M$ tends to the cutoff of the theory (so that $V_{NN}\sim M^2$) we can neglect the temporal derivatives in the expansion above relative to the mass term and the spatial derivatives (which always become significant at horizon crossing), thus reducing the Green's function to leading order to only the contact term \footnote{A related derivation for the effective field theory of the inflaton field coupled to a massive field with a cubic interaction term with the inflaton can be found in \cite{rubin}. We credit \cite{dong} in bringing this reference to our attention.}. Integrating the second term in (\ref{effa}) by parts results in:
\begin{equation}
S =  \int~d\tau d^3k~\frac{1}{2} \left\{ \left(\frac{d v^T}{d\tau}\right)^2 e^{\beta(k , \tau)} -   \left[ k^2  + \bar \Omega (\tau,k)  \right] \left(v^T\right)^2 \right\} \, ,
\end{equation}
with
\begin{align}
e^{\beta(\tau, k^2)} \equiv & \, 1+ {4 \eta_{\bot}^2} \left( \frac{M^2}{H^2} - 2 + \epsilon - \eta_{\bot}^2 + \frac{k^2}{a^2H^{2}} \right)^{-1} \, ,
\\
\bar \Omega(\tau,k) \equiv & \,  \Omega_{0}  -  \frac{4 a^4 H^4 \eta_{\bot}^2 (1 + \epsilon - \eta_{||})^2 }{\omega^2}  - 4 \frac{d}{d \tau} \left[  \frac{ a^3 H^3 \eta_{\bot}^2 (1 + \epsilon - \eta_{||})}{\omega^2} \right] \, ,
\label{betadef2}
\end{align}
$\Omega_0$ is given by (\ref{omega-0}). Making the field redefinition $\varphi \equiv e^{\beta/2}v^T$ (and upon integrating by parts the resulting friction term), after some manipulations one then obtains the effective action
\begin{equation}
S =  \int~d\tau d^3k~\frac{1}{2}\left[ \left(\frac{d \varphi}{d\tau}\right)^2 - \varphi~ e^{-\beta(\tau,k)} k^2\varphi - \varphi~\Omega(\tau,k) \varphi \right] ,
\end{equation}
where $\Omega(\tau , k^2)$ is defined as in (\ref{omega-k}). Thus we see that expression (\ref{action-effective}) follows.

\end{appendix}

\end{document}